\documentclass[11pt]{article}%
\usepackage{amssymb}
\usepackage{amsfonts}
\usepackage{amsmath}
\usepackage{amssymb}
\usepackage{color}
\usepackage{graphicx}
\usepackage{caption}%
\usepackage{hyperref}
\usepackage{float}
\usepackage{subcaption}
\usepackage{academicons}

\usepackage{tabularx}

\providecommand{\U}[1]{\protect\rule{.1in}{.1in}}
\newtheorem{theorem}{Result}

\newtheorem{remark}[theorem]{Remark}
\newtheorem{definition}[theorem]{Definition}
\newenvironment{proof}[1][Proof]{\noindent \textbf{#1.} }{\hspace{16cm} \rule{0.5em}{0.5em}}
\newcommand\blfootnote[1]{%
	\begingroup
	\renewcommand\thefootnote{}\footnote{#1}%
	\addtocounter{footnote}{-1}%
	\endgroup
}

\begin{document}
	
	\title{\textbf{Robust inference for non-destructive one-shot device testing under step-stress model with exponential lifetimes  }}
	\author{Narayanaswamy Balakrishnan$^1$, Elena Castilla$^2$, Mar\'ia Jaenada$^2$ and Leandro Pardo$^2$}
	\date{ }
	\maketitle
	
$^1$Department of Mathematics and Statistics, McMaster University, Hamilton, Ontario, Canada.\\
\indent $^2$Department of Statistics and O.R., Complutense University of Madrid, Madrid, Spain.\\

\blfootnote{Correspondence: N. Balakrishnan: bala@mcmaster.ca; E. Castilla: elecasti@ucm.es; M.Jaenada: mjaenada@ucm.es	and L. Pardo: lpardo@mat.ucm.es}

\begin{abstract}
		%IDEAS: ALT for reliability. Non-destructive devices still give information. SSALT under exponential lifetimes. Robust estimators and Z-type test. 
	
	One-shot devices analysis involves an extreme case of interval censoring, wherein one can only know whether the failure time is either before or after the test time. 
	Some kind of one-shot devices do not get destroyed when tested, and so can continue within the experiment, providing extra information for inference, if they did not fail before an inspection time. 
	In addition, their reliability can be rapidly estimated via accelerated life tests (ALTs) by running the tests at varying and higher stress levels than working conditions. In particular, step-stress tests allow the experimenter to increase the stress levels at pre-fixed times gradually during the life-testing experiment.  
	The cumulative exposure model is commonly assumed for step-stress models, relating the lifetime distribution of units at one stress level to the lifetime distributions at preceding stress levels. 
	%Besides, robust methods have recently shown to offer benefits in one-shot devices inference, offering a competetive alternative to the classical maximum likehood estimator.
	In this paper,
	%we assume exponential lifetime distributions for the one shot devices, and 
	we develop robust estimators and Z-type test statistics based on the density power divergence (DPD) for testing linear null hypothesis for non-destructive one-shot devices under the step-stress ALTs with exponential lifetime distribution. We study asymptotic and robustness properties of the estimators and test statistics, yielding point estimation and confidence intervals for different lifetime characteristic such as reliability, distribution quantiles and mean lifetime of the devices. A simulation study is carried out to assess the performance of the methods of inference developed here and some real-life data sets are analyzed finally for illustrative purpose. 
	
	%The study of one-shot device testing has been developed considerably recently, both 	in terms of estimation and optimal design under different lifetime distributions.

	%A reliability experimenter is often interested in studying the effects of extreme or varying stress factors such as load, pressure, temperature and voltage on the lifetimes of experimental units.  
	%Finally, we present some examples to illustrate	all the inferential methods discussed here.
\end{abstract}
\section{Introduction}

One-shot device testing is an increasingly important problem in the area of reliability. This involves an extreme case of interval censoring, wherein one only knows if the device works when it is tested. Most of the existing literature considers the case of ``destructive'' one-shot devices. This is the case when, once the device is used, it is either destroyed or must be rebuilt. Some typical examples are automobile air bags, fuel injectors, disposable napkins,  missiles (Olwell and Sorell, 2001) and fire extinguishers and munition (Newby, 2008).  Mainly motivated by the work of  Fan et al. (2009), Balakrishnan and Ling (2012, 2013, 2014) developed efficient EM algorithms for the estimation of model parameters under the assumption of exponential, Weibull and gamma lifetime distributions, respectively. One may refer to the recent book by Balakrishnan et al. (2021) for a detailed
review of all these works.  Balakrishnan and Castilla (2021) recently developed  results for the  lognormal lifetime distribution. Some other related works about  one-shot devices can be found in  Mun et al. (2013), Sharma and Upadhyay (2018) and Zhu  et al. (2021). However, the destructiveness assumption is not always necessary as in many experiments, the tested devices can be reused if has not failed during the test. We will refer to this type of  devices as ``non-destructive'' one-shot devices.
Their major advantage is that operating devices can continue in the experiment, providing extra information about their lifetime characteristics.
 Some typical  examples are metal fatigue, thermal ageing of electrical insulation, spare wheels, safety valves, hot spare disks, electronics components, light bulbs, 
 %electric hand drills, 
 electric motors and
 %and bearing balls.
stability of pharmaceuticals.

A common practice in reliability is to employ accelerated life tests (ALTs) to shorten the lifetime of a product by increasing some stress factors associated with it, such as temperature, pressure or humidity. This way, the experimental time and cost can be reduced. After suitable inference is developed, we can then extrapolate the results to normal operating conditions; see Meeter and Mecker (1994) and Meeker et al. (1998). 
There are different types of ALTs resulting in different statistical models. For example, constant-stress ALTs  assume that each device is subject to only pre-specified stress levels, while step-stress ALTs  apply stress to devices in such a way they will get changed at pre-specified times, and progressive-stress ALT continuously increases the stress level. The constant-stress and step-stress ALTs have been widely studied for destructive one-shot devices; see  Ling (2019), Lee and Bae (2020), Wu et al. (2020) and Ling and Hu (2020), among others. We  focus here on the step-stress model for non-destructive one-shot devices. In particular, we adopt a parametric approach in which the  lifetimes of the devices are assumed to follow exponential distribution.

While classical estimation methods are based on the maximum likelihood estimators (MLE), recent works have shown the advantage of using divergence-based methods in terms of robustness, with an unavoidable loss of efficiency in the case of uncontaminated data. Balakrishnan et al. (2019a, 2019b, 2020a, 2020b, 2021) developed robust estimation methods based on density power divergence (DPD) for the constant-stress model and then constructed robust test statistics for testing linear hypothesis.
%In this line, most of the literature on robust methods is focused on constant-ALT models. 
 In this paper, we develop robust estimators and test statistics for non-destructive one-shot devices under the multiple step-stress model and exponential lifetimes, and illustrate their robustness features both theoretically and empirically. 

The rest of the paper is organized as follows: Section \ref{sec:modelformulation} describes the multiple step-stress accelerated life test (SSALT) under exponential lifetimes and introduces the classical MLE for the SSALT model and Section \ref{sec:MDPDE} presents the minimum DPD and the minimum restricted DPD estimators under linear constraints, with the corresponding asymptotic results. In Section \ref{sec:IF} the robustness of the proposed estimators is examined through their influence function analysis. Section \ref{sec:interval} describes point estimation and confidence intervals for the reliability, distributional quantile and mean lifetime of the device based on minimum DPD estimators. In Section \ref{sec:robusttest}, robust test statistics are developed, including Z-type and Rao-type tests, and their asymptotic properties are examined, providing approximate power functions of the tests. Sections \ref{sec:simstudy} and \ref{sec:realdata} empirically illustrate the performance of the proposed methods thorough an extensive simulation study and real data analysis respectively. Finally, Section \ref{sec:concluding} presents some concluding remarks.

\section{Model formulation and the maximum likelihood estimator} \label{sec:modelformulation}

Let us consider a SSALT with $k$ ordered stress levels, $ x_1 < x_2 < \dots < x_k,$ and $N$ one-shot devices under test. 
At pre-fixed times $\tau_i,$ called times of stress change,  we increase the stress level from $x_{i}$ to $x_{i+1},$ for $i=1,...,k-1,$ and we denote $\tau_k$ the time at which the experiment terminates.
Let us also consider a sequence of length $L$ of inspection times during the experiment, including the times of stress change $\tau_i,$ $i=1,...,k$, 
$$0 < t_1 < \dots < t_{l_1} = \tau_1 < t_{l_1+1} < \dots < t_{l_k} = \tau_k, $$
where $l_i$ denotes the number of inspection times before the $i$-th stress change and $L = l_k.$ Under this set-up, the simple step-stress model corresponds to the case when $k=2,$ $l_1= 1$ and $l_2 =2.$ This model has been widely studied in the literature; for example, Nelson (1980) discussed general cumulative exposure model, including the simple stress model, while Balakrishnan (2009) reviewed exact inferential procedures for exponential step-stress models.

%Sedyakin (1966), Bagdonavicius (1978) and Nelson (1980) all studied the cumulative exposure model and associated inference. 

%Madi (1993) generalized this simple tampered failure rate model to the multiple step-stress case. 
 
%Xiong (1998) and Xiong and Milliken (1999) discussed inference for the exponential step-stress model by assuming that the mean lifetime of the experimental units at the ith stress level si has a log-linear form. 

%Gouno et al. (2004) and Han et al. (2006) discussed inferential methods for step-stress models under the exponential distribution when the available data are progressively Type-I censored. 

%Xiong and Ji (2004) discussed the analysis of step-stress life-tests based on grouped and censored data. While these discussions all focused on inference for exponential step-stress models, Khamis and Higgins (1998) and Kateri and Balakrishnan (2008) examined inferential methods for a cumulative exposure model with Weibull distributed lifetimes. 

%Comprehensive reviews of work on step-stress models have been provided by Gouno and Balakrishnan (2001) and Tang (2003).

We further assume that the lifetime, $T,$ of a device follows an exponential distribution, under stress level $x_i,$  with failure rate $\lambda_i$ depending on the stress.
The distribution of the lifetime of a device during the test is then formed under applying the cumulative exposure model, which relates the lifetime distribution of a device at one stress level to the distributions at preceding stress levels by assuming the residual life of that device depends only on the cumulative exposure it had experienced, with no memory of how this exposure was accumulated. Then, if $G_i(\cdot)$ denotes the exponential lifetime distribution function at the $i$-th stress level, the distribution function of $T,$ $G_T(t),$ is given by
\begin{equation}\label{eq:distributionT}
	G_T(t) = 
	\begin{cases}
		G_1(t) = 1-e^{-\lambda_1 t}, & 0<t< \tau_1\\
		G_2\left(t + a_1 -\tau_1 \right)  = 1-e^{-\lambda_2(t + a_1 -\tau_1)}, & \tau_1 \leq t < \tau_2 \\
		\vdots  & \vdots \\
		G_k\left(t + a_{k-1} -\tau_{k-1} \right)  =  1-e^{-\lambda_k(t + a_{k-1} -\tau_{k-1})}, & \tau_{k-1} \leq t < \infty ,\\
	\end{cases}
\end{equation}
with 
\begin{equation}\label{eq:ai}
	a_{i-1} = \frac{ \sum_{l=1}^{i-1}\left(\tau_l-\tau_{l-1}\right)\lambda_l}{\lambda_i},
\end{equation}
 for $i=1,...,k-1.$ For notational convenience, we set $a_{-1} = \tau_{-1}=0.$
The corresponding density function of $T$ is given by
\begin{equation} \label{eq:densityT}
	g_T(t) = 
	\begin{cases}
		g_1(t) = \lambda_1 e^{-\lambda_1 t}, & 0<t <\tau_1\\
		g_2\left(t + a_1 -\tau_1 \right)  = \lambda_2e^{-\lambda_2(t + a_1 -\tau_1)}, & \tau_1 \leq t < \tau_2 \\
		\vdots & \vdots \\
		g_k\left(t + a_{k-1} -\tau_{k-1} \right)  =  \lambda_ke^{-\lambda_k(t + a_{k-1}, -\tau_{k-1})} & \tau_{k-1} \leq t < \infty .\\
	\end{cases}
\end{equation}
%It is not a continuous function!!!!!!!!!!
Although the distribution function is continuos in $(0,\infty),$ the density function has $k$  points of discontinuity at times of stress change. 
We further assume that at  stress level $x_i$, the rate parameter $\lambda_i$ of a device  has a log-linear relationship with stress level given by
\begin{equation} \label{eq:loglinear}
	\lambda_i(\boldsymbol{\theta}) = \theta_0 \exp(\theta_1 x_i), \hspace{0.3cm} i = 1,..,k,
\end{equation} 
where $\boldsymbol{\theta} = (\theta_0, \theta_1) \in \mathbb{R}^+ \times \mathbb{R} = \Theta$ is an unknown parameter vector of the model. Note that the mean  lifetime of a device is inverse of the exponential parameter, and so it would decrease with an increase in the level.
The log-linear relation in (\ref{eq:loglinear}) is frequently assumed in accelerated life test models, as it can be shown to be equivalent to  the well-known inverse power law model or the Arrhenius reaction rate model.

% We assume that there are units being tested at every stage of stress change
%why the loglinear link? explain

%\section{The maximum likelihood estimator}\label{sec:MLE}
%%%%%
%When the SSALT is carried out, the number of failures at each interval $(t_{j-1}, t_j],$ $n_j,$ is recorded, $j=1,..,L$. 
Suppose $n_j$ failures of test devices are observed %while testing at stress $x_i$ 
in the interval $(t_{j-1}, t_j],$ $j=1,..,k,$ and
for notational ease, let us denote $n_{L+1}$  for the number of surviving devices at the end of the experiment. 
Then, the probability of failure of a device in the $j$-th interval is
\begin{equation} \label{eq:th.prob}
		\pi_j(\boldsymbol{\theta}) =  G_T(t_j) - G_T(t_{j-1}), \hspace{0.3cm} j = 1,..,L,
\end{equation}
and the probability of survival at the end of the experiment is $\pi_{L+1}(\boldsymbol{\theta}) = 1 - G_T(t_L).$
Accordingly, a multinomial model with probability vector $\boldsymbol{\pi} (\boldsymbol{\theta}) = (\pi_1(\boldsymbol{\theta})),...,\pi_{L+1}(\boldsymbol{\theta}))^T$ and $N$ trials can be used to present the likelihood function of the model as
$$\mathcal{L}(\boldsymbol{\theta}; n_1,..,n_{L+1}) = \frac{N!}{n_1!\cdots n_{L+1}!} \prod_{j=1}^{L+1} \pi_j(\boldsymbol{\theta})^{n_j}.$$
From the above likelihood function, the MLE  of $\boldsymbol{\theta}$ would simply be 
$$
	\boldsymbol{\widehat{\theta}}^{MLE} = \left(\widehat{\theta}_0^{MLE}, \widehat{\theta}_1^{MLE}\right) = \operatorname{arg} \operatorname{max}_{\boldsymbol{\theta}  \in \Theta} \mathcal{L}(\boldsymbol{\theta}; n_1,..,n_{L+1}).
$$

\begin{remark}
	We could have alternatively derived the likelihood function of the model using binomial distribution for each interval, by using conditional probabilities of failure, given that the device did not fail in earlier time intervals. However, both approaches yield  the same likelihood function.
\end{remark}

Now, 
%$N_j$ denote the number of survival units at the beginning of the $j$-th interval, and 
let $\widehat{\boldsymbol{p}} = \left(n_1/N, ..., n_{L+1}/N\right)$ be the empirical probability vector obtained from the observed data. Then, the Kullback-Leibler divergence between the empirical and theoretical probability vectors, $\widehat{\boldsymbol{p}}$ and $\boldsymbol{\pi} (\boldsymbol{\theta}),$ is given by
$$d_{KL} (\boldsymbol{\widehat{p}}, \boldsymbol{\pi}(\boldsymbol{\theta})) = \sum_{j=1}^{L+1}\widehat{p}_j \log\left(\frac{\widehat{p}_j}{\pi_j(\boldsymbol{\theta})}\right).$$
It is straightforward to see that the Kullback-Leibler divergence is related to the log likelihood function in the form
\begin{equation}\label{eq:KL}
	d_{KL} (\boldsymbol{\widehat{p}}, \boldsymbol{\pi}(\boldsymbol{\theta}))  = c - \frac{1}{N}\log \mathcal{L}(\boldsymbol{\theta}; n_1,..,n_{L+1})
\end{equation}
where the constant $c = \sum_{i=1}^{L+1} \widehat{p}_i\log(\widehat{p}_i)$ does not depend on $\boldsymbol{\theta}.$ Hence, the MLE can equivalently be defined as
\begin{equation} \label{eq:MLE}
	\boldsymbol{\widehat{\theta}}^{MLE} = \operatorname{arg} \operatorname{min}_{\boldsymbol{\theta}  \in \Theta} d_{KL} (\boldsymbol{\widehat{p}}, \boldsymbol{\pi}(\boldsymbol{\theta})).
\end{equation}

From an asymptotic point of view, it is well-known that the MLE is a BAN (Best Asymptotically Normal) estimator, and it has therefore been widely used for the SSALT model.
However, despite its high efficiency, the MLE lacks robustness as contaminated data could influence the parameter estimation considerably. 
In the next section, we present a robust family of estimators for the SSALT model based on the DPD.

\section{Minimum density power divergence estimator}\label{sec:MDPDE}

The density power divergence (DPD) family, introduced by Basu et al (1998), a rich class of density-based divergences,  produces robust estimators with relative small loss in efficiency. Given two density or mass functions, $f_{\boldsymbol{\theta}}$ and $g,$ the DPD between them is defined as
\begin{equation*}\label{eq:DPD}
	d_{\beta}(g, f_{\boldsymbol{\theta}}) =\int \left\{ f_{\boldsymbol{\theta}}^{1+\beta}(y)-\frac{\beta +1}{\beta }f_{\boldsymbol{\theta}}^{\beta}(y)g(y)+\frac{1}{\beta}g^{1+\beta}(y)\right\} dy \hspace{0.5cm} \text{for } \beta >0.
\end{equation*}
The parameter $\beta,$ indexing the DPD divergence, controls the trade-off between efficiency and robustness. In fact, the DPD can be defined at $\beta = 0$ by taking continuous limits leading to the Kullback-Leibler divergence. 

Following the discussion in the last section, we consider the DPD between the empirical and theoretical probability vectors, $\widehat{\boldsymbol{p}}$ and $\boldsymbol{\pi} (\boldsymbol{\theta}),$ 
\begin{equation}\label{eq:DPDloss}
	d_{\beta}\left( \widehat{\boldsymbol{p}},\boldsymbol{\pi}\left(\boldsymbol{\theta}\right)\right)   = \sum_{j=1}^{L+1} \left(\pi_j(\boldsymbol{\theta})^{1+\beta} -\left( 1+\frac{1}{\beta}\right) \widehat{p}_j\pi_j(\boldsymbol{\theta})^{\beta}  +\frac{1}{\beta} \widehat{p}_j^{\beta+1} \right),
\end{equation}
and correspondingly define the minimum DPD estimator (MDPPE) as
\begin{equation}
	\boldsymbol{\widehat{\theta}}^{\beta} = \left(\widehat{\theta}_0^{\beta},\widehat{\theta}_{1}^{\beta}\right) = \operatorname{arg} \operatorname{min}_{\boldsymbol{\theta}  \in \Theta} d_{\beta} \left( \widehat{\boldsymbol{p}},\boldsymbol{\pi}\left(\boldsymbol{\theta}\right)\right).
\end{equation}
Note that the value $\beta = 0$ corresponds to the MLE of $\boldsymbol{\theta}$. Hence, the proposed family could be considered as a generalization of the MLE with a tuning parameter $\beta$ accounting for the compromise between efficiency and robustness. 
Moreover, the last term of each addend in (\ref{eq:DPDloss}) does not depend on the model parameter, and so it can be ignored in the minimization process.

The next result presents the estimating equations for the MDPDE.
\begin{theorem}\label{thm:esitmatingequations}
	The estimating equations associated with the MDPDE for the SSALT model, under exponential lifetimes, satisfying the log-linear relation in (\ref{eq:loglinear}), are given by
	$$\boldsymbol{W}^T\boldsymbol{D}_{\boldsymbol{\pi}(\boldsymbol{\theta})}^{\beta-1}\left( \widehat{\boldsymbol{p}}- \boldsymbol{\pi}(\boldsymbol{\theta})\right) = \boldsymbol{0}_2,$$
	where $\boldsymbol{0}_2$ is the 2-dimensional null vector, $\boldsymbol{D}_{\boldsymbol{\pi}(\boldsymbol{\theta})}$ denotes a $(L+1)\times(L+1)$ diagonal matrix with diagonal entries $\pi_j(\boldsymbol{\theta}),$ $j=1,...,L+1,$ and $\boldsymbol{W}$ is a $(L+1) \times 2$ matrix with rows
	$
		\boldsymbol{w}_j =  \boldsymbol{z}_j-\boldsymbol{z}_{j-1},
	$
	where
	\begin{align}
		\label{eq:zj} \boldsymbol{z}_j  &= g_T(t_j)\begin{pmatrix}
			\frac{t_j+a_{i-1}-\tau_{i-1}}{\theta_0}\\
			(t_j+a_{i-1}-\tau_{i-1})x_i + a_{i-1}^\ast
		\end{pmatrix}, \hspace{0.3cm} j = 1,...,L,\\ 
	a_{i-1}^\ast &= \frac{1}{\lambda_{i}} \sum_{l=1}^{i-1}\lambda_l\left(\tau_l-\tau_{l-1}\right)(-x_{i}+x_l) , \hspace{0.3cm} i=2,..,k,\label{aast}
	\end{align}
 $ \boldsymbol{z}_{-1} = \boldsymbol{z}_{L+1} = \boldsymbol{0}$ and $i$ is the stress level at which the units are tested after the $j-$th inspection time.
\end{theorem}

For the MLE, the estimating equations are obtained by deriving the Kullback-Leibler divergence given in (\ref{eq:KL}), yielding

$$\boldsymbol{W}^T\boldsymbol{D}_{\boldsymbol{\pi}(\boldsymbol{\theta})}^{-1}\left( \widehat{\boldsymbol{p}}- \boldsymbol{\pi}(\boldsymbol{\theta})\right) = \boldsymbol{0}_2.$$

Next we present the asymptotic distribution of the proposed estimator, for any positive value of $\beta$.
\begin{theorem} \label{thm:asymptoticestimator}
	Let $\boldsymbol{\theta}_0$ be the true value of the parameter $\boldsymbol{\theta}$. Then, the asymptotic distribution of the MDPDE, $\boldsymbol{\widehat{\theta}}^{\beta},$ for the SSALT model, under exponential lifetime, is given by
	$$ \sqrt{N}\left(\boldsymbol{\widehat{\theta}}^{\beta} - \boldsymbol{\theta}_0\right) \rightarrow \mathcal{N}\left(\boldsymbol{0}, \boldsymbol{J}_\beta^{-1}(\boldsymbol{\theta}_0)\boldsymbol{K}_\beta(\boldsymbol{\theta}_0)\boldsymbol{J}_\beta^{-1}(\boldsymbol{\theta}_0)\right),$$
	where \begin{equation} \label{eq:JK}
		\boldsymbol{J}_\beta(\boldsymbol{\theta}_0) = \boldsymbol{W}^T D_{\boldsymbol{\pi}(\boldsymbol{\theta_0})}^{\beta-1} \boldsymbol{W},
		\hspace{0.3cm}  \hspace{0.3cm}
		\boldsymbol{K}_\beta(\boldsymbol{\theta}_0) = \boldsymbol{W}^T \left(D_{\boldsymbol{\pi}(\boldsymbol{\theta_0})}^{2\beta-1}-\boldsymbol{\pi}(\boldsymbol{\theta}_0)^{\beta}\boldsymbol{\pi}(\boldsymbol{\theta}_0)^{\beta T}\right) \boldsymbol{W},
	\end{equation}
 $D_{\boldsymbol{\pi}(\boldsymbol{\theta_0})}$ denotes the diagonal matrix with entries $\pi_j(\boldsymbol{\theta_0}),$ $j=1,...,L+1,$ and $\boldsymbol{\pi}(\boldsymbol{\theta}_0)^{\beta}$ denotes the vector with components $\pi_j(\boldsymbol{\theta}_0)^{\beta}.$
\end{theorem}

For $\beta = 0$, the Fisher information matrix associated with the SSALT model under exponential lifetimes coincides with the matrices $\boldsymbol{J}_\beta(\boldsymbol{\theta}_0)$ and $\boldsymbol{K}_\beta(\boldsymbol{\theta}_0),$ and so we obtain the asymptotic distribution of the MLE as a particular case, i.e.,
$$ \sqrt{N}\left(\boldsymbol{\widehat{\theta}}^{0} - \boldsymbol{\theta}_0\right) \rightarrow \mathcal{N}\left(\boldsymbol{0}, \boldsymbol{I}_F^{-1}(\boldsymbol{\theta}_0)\right),$$
where $\boldsymbol{I}_F(\boldsymbol{\theta}_0) =  \boldsymbol{W}^T D_{\boldsymbol{\pi}(\boldsymbol{\theta_0})}^{-1} \boldsymbol{W}.$ 
%Las matrices J y K para beta =0 son las matrices de informacion de fisher del modelo habría que ponerlas como un corolario o comntario e indicar si esa matriz o la correspondiente matriz para un modelo mas sencillo que el nuestro se encuentra en algún artículo y citar rlarticulo

\begin{remark}\label{remark:4}
As $\widehat{\boldsymbol{\theta}}^\beta$ is a consistent estimator of $\boldsymbol{\theta}_0$, the asymptotic variances of $\widehat{\theta}_0^\beta$ and $\widehat{\theta}_1^\beta$ for $\beta >0,$  denoted by $\sigma^2(\theta_0^\beta)$ and $\sigma^2(\theta_1^\beta),$ respectively, can be estimated by the diagonal entries of $\boldsymbol{J}_\beta^{-1}(\widehat{\boldsymbol{\theta}}^\beta)\boldsymbol{K}_\beta(\widehat{\boldsymbol{\theta}}^\beta)\boldsymbol{J}_\beta^{-1}(\widehat{\boldsymbol{\theta}}^\beta)$.
Therefore, asymptotic confidence intervals for $\theta_0$ and $\theta_1,$  at  confidence level $(1-\alpha),$ are given by 
\begin{equation}\label{eq:CI}
	\widehat{\theta}_i^\beta \pm \frac{\widehat{\sigma}(\theta_i^\beta)}{\sqrt{N}}z_{\alpha/2}, \hspace{0.3cm} i = 0,1,
\end{equation} 
with $z_{\alpha/2}$ being the lower $\alpha/2$-quantile of a standard normal distribution.
Moreover, we have 
$$N(\widehat{\boldsymbol{\theta}}^\beta - \boldsymbol{\theta})^T(\boldsymbol{J}_\beta^{-1}(\boldsymbol{\theta}_0)\boldsymbol{K}_\beta(\boldsymbol{\theta}_0)\boldsymbol{J}_\beta^{-1}(\boldsymbol{\theta}_0))^{-1}(\widehat{\boldsymbol{\theta}}^\beta - \boldsymbol{\theta}) \rightarrow \chi_2^2$$
and consequently an associated ellipsoidal confidence region for $\boldsymbol{\theta} = (\theta_0,\theta_1)$ is given by 
$$C_{N,\beta}^\alpha = \{\theta | N(\widehat{\boldsymbol{\theta}}^\beta - \boldsymbol{\theta})^T(\boldsymbol{J}_\beta^{-1}(\boldsymbol{\theta}_0)\boldsymbol{K}_\beta(\boldsymbol{\theta}_0)\boldsymbol{J}_\beta^{-1}(\boldsymbol{\theta}_0))^{-1}(\widehat{\boldsymbol{\theta}}^\beta - \boldsymbol{\theta}) \leq c\}.$$
Choosing $c=\chi_{2,\alpha}^2$, the $100(1-\alpha)$-percentile of the chi-square distribution with 2 degrees of freedom, 
%i.e., $\mathbb{P}(\chi_2^2 > \chi_{2,\alpha}^2) = \alpha,$ 
we have  $\mathbb{P}_{\boldsymbol{\theta}}(C_{N,\beta}^\alpha)$ tending to $1-\alpha$ as $n \rightarrow \infty$. Hence, $C_{n,\beta}^\alpha$ represents an ellipsoidal confidence region for $\boldsymbol{\theta}$ having limiting confidence coefficient $1-\alpha$ as $n\rightarrow \infty$ (see Serfling (2009) for more details). The volume of the ellipsoidal region $C_{N,\beta}^\alpha,$ based on Cramer (1946), is given by
$$\chi_{2,\alpha}^2 \pi  \operatorname{det}\left(\left(\boldsymbol{J}_\beta^{-1}(\widehat{\boldsymbol{\theta}}^\beta)\boldsymbol{K}_\beta(\widehat{\boldsymbol{\theta}}^\beta)\boldsymbol{J}_\beta^{-1}(\widehat{\boldsymbol{\theta}}^\beta) \right)^{-1}\right)^{1/2}. $$
%so we can compare the volume of $C_{n,\beta}^\alpha,$ $\beta > 0$ with respect to the volume of  $C_{n,\beta=0}^\alpha$ we can get that 
Thus, a measure of the asymptotic relative efficiency of $\widehat{\boldsymbol{\theta}}^\beta,$ for $\beta >0,$ with respect to the MLE, $\widehat{\boldsymbol{\theta}}^0,$ is given by
$$\left(\frac{\operatorname{det}\left(\boldsymbol{I}_F(\widehat{\boldsymbol{\theta}})\right)}{|\boldsymbol{J}_\beta^{-1}(\widehat{\boldsymbol{\theta}}^\beta)\boldsymbol{K}_\beta(\widehat{\boldsymbol{\theta}}^\beta)\boldsymbol{J}_\beta^{-1}(\widehat{\boldsymbol{\theta}}^\beta)|}\right)^{1/2}.$$
\end{remark}

\section{Influence function of the MDPDE} \label{sec:IF}

The influence function (IF), first introduced by Hampel et al. (1986), plays a central role in the study of  robustness properties of an estimator. Intuitively, it quantifies the impact of an infinitesimal perturbation in the true distribution underlying the data on the asymptotic value of the resulting parameter estimate. An estimator is said to be robust if its influence function is bounded.

%estimator viewed as a functional
Mathematically, the IF of an estimator is computed in terms of its corresponding statistical functional. 
%For the MDPDE of the SSALT model, $\widehat{\boldsymbol{\theta}}^\beta,$ 
Let $F_{\boldsymbol{\theta}}$ and $\boldsymbol{G}$ be the assumed distribution of the model and the true density underlying the data, respectively. We use $\boldsymbol{T}(\boldsymbol{G})$ to denote the statistical functional associated with the estimator $\widehat{\boldsymbol{\theta}}.$
Then, the IF of the estimator $\widehat{\boldsymbol{\theta}}$ at a point $\boldsymbol{t}$ is computed as 
\begin{equation} \label{eq:IFmath}
	\text{IF}\left(\boldsymbol{t}, \boldsymbol{T}, \boldsymbol{G}\right) = \lim_{\varepsilon \rightarrow 0}\frac{\boldsymbol{T}(\boldsymbol{G}_\varepsilon)- \boldsymbol{T}(\boldsymbol{G})}{\varepsilon} = \frac{\partial \boldsymbol{T}(G_\varepsilon)}{\partial \varepsilon}\bigg|_{\varepsilon = 0},
\end{equation} 
where $\boldsymbol{G}_\varepsilon = (1-\varepsilon)\boldsymbol{G} + \varepsilon\Delta_{\boldsymbol{t}}$ is the contaminated version of $\boldsymbol{G},$ with $\varepsilon$ being the contamination proportion, and $\Delta_{\boldsymbol{t}}$ being the degenerate distribution at the contamination point $\boldsymbol{t}.$ 
An estimator is said to be robust if its IF at the model distribution $F_{\boldsymbol{\theta}}$ is bounded.
For the SSALT model, we could consider only one cell contamination, and so the contamination point $ \boldsymbol{t}$ would have all elements equal to zero except for only one component. 

%\subsection{Influence function of the minimum density power divergence estimator}\label{sec:IFMDPDE}

Let us denote $F_{\boldsymbol{\theta}}$ for the assumed distribution of the multinomial model with mass function $\boldsymbol{\pi}(\boldsymbol{\theta})$ given by the SSALT model with exponential lifetimes
and $\boldsymbol{G}$ denote the true distribution underlying the data, with mass function $\boldsymbol{g}.$ 
%\begin{equation}\label{eq:H}
%	H_{\beta}\left( g,\boldsymbol{\pi}\left(\boldsymbol{\theta}\right)\right)   = \sum_{j=1}^{L+1} \left(\pi_j(\boldsymbol{\theta})^{1+\beta} -\left( 1+\frac{1}{\beta}\right) g_j\pi_j(\boldsymbol{\theta})^{\beta} + \frac{1}{\beta}g_j^{1+\beta} \right)
%\end{equation}
We define the statistical functional $\boldsymbol{T}_\beta(\boldsymbol{G})$ as the minimizer of the DPD between the two mass functions, $\boldsymbol{\pi}_{\boldsymbol{\theta}}$ and $\boldsymbol{g},$  given in (\ref{eq:DPD}). Then, an expression of the IF can be computed from (\ref{eq:IFmath}) as stated in the following result.
\begin{theorem}\label{thm:IF}
	The IF of the MDPDE of the SSALT model, $\widehat{\boldsymbol{\theta}}^\beta,$ at a point contamination $\boldsymbol{n}$ and the assumed model distribution $F_{\boldsymbol{\theta}_0}$ is given by
	\begin{equation}\label{eq:IF}
		\text{IF}\left(\boldsymbol{n}, \boldsymbol{T}_\beta, F_{\boldsymbol{\theta}_0} \right) = \boldsymbol{J}_\beta^{-1}(\boldsymbol{\theta}_0) \boldsymbol{W}^T \boldsymbol{D}_{\boldsymbol{\pi}(\boldsymbol{\theta}_0)}^{\beta-1}\left(-\boldsymbol{\pi}(\boldsymbol{\theta}_0)+\Delta_{\boldsymbol{n}} \right).
	\end{equation}
\end{theorem}

\begin{remark}
The matrix $\boldsymbol{J}_\beta(\boldsymbol{\theta}_0)$ is assumed to be bounded, and so the robustness of the estimators depends on the boundedness of the second factor of the IF, given by
\begin{equation}\label{eq:secondfactorIF}
	\boldsymbol{W}^T \boldsymbol{D}_{\boldsymbol{\pi}(\boldsymbol{\theta}_0)}^{\beta-1}\left(-\boldsymbol{\pi}(\boldsymbol{\theta}_0)+\Delta_{\boldsymbol{n}} \right) = \sum_{j=1}^{L+1} (\boldsymbol{z}_j -\boldsymbol{z}_{j-1} )\pi_j(\boldsymbol{\theta}_0)^{\beta-1} \left(-\pi_j(\boldsymbol{\theta}_0)+
	 \Delta_{\boldsymbol{n}j}\right),
	\end{equation}
where $\boldsymbol{z}_j$ is as defined in (\ref{eq:zj}). 
All the terms in  (\ref{eq:secondfactorIF}) are bounded for fixed stress levels and inspection times at any contamination point $\boldsymbol{n},$ as the maximum value of its components is the number of trials $N.$ Then, any of the proposed MDPDE for $\beta \geq 0$ is robust against vertical outliers, including the MLE.
Conversely, the IF boundedness is affected by leverage points, i.e. outlier inspection times or outlier stress levels.
% whereas the MDPDEs for positive values of $\beta$ remain robust. 

Let us first consider the situation wherein an inspection time, $t_j,$ tends to infinity, for fixed $j$. We denote $i$ the fixed stress level corresponding to the $j-$th inspection time. 
As the inspection times are ordered, there will be no more terms in the summation after the $j$-th term.
%We must establish the boundedness of all terms of the summation from the $j-$th onwards. Since the inspection times are ordered, we have that $t_l \rightarrow \infty$ for all $l \geq j.$ 
%For the $j$-th term, 
Then, we can write
\begin{align*}
	(\boldsymbol{z}_j-\boldsymbol{z}_{j-1})\pi_j(\boldsymbol{\theta}_0)^{\beta-1} =& 
	\left( g_i(T_j )\begin{pmatrix}
		\frac{T_j}{\theta_0}\\
		T_jx_i + a_{i-1}^\ast
	\end{pmatrix} - g_i(T_{j-1} )\begin{pmatrix}
		\frac{T_{j-1}}{\theta_0}\\
		T_{j-1}x_i + a_{i-1}^\ast
	\end{pmatrix} 
	\right)(G_i(T_j)-G_i(T_{j-1}))^{\beta-1}\\
	=& 
	\bigg[ \lambda_i(\boldsymbol{\theta}_0)\exp(- \lambda_i(\boldsymbol{\theta}_0) T_j)\begin{pmatrix}
		\frac{T_j}{\theta_0}\\
		T_jx_i + a_{i-1}^\ast
	\end{pmatrix}  
	-  \lambda_i(\boldsymbol{\theta}_0) \exp(- \lambda_i(\boldsymbol{\theta}_0)T_{j-1})\begin{pmatrix}
		\frac{T_{j-1}}{\theta_0}\\
		T_{j-1}x_i + a_{i-1}^\ast
	\end{pmatrix} 
	\bigg]\\
	& \left( - \exp( - \lambda_i(\boldsymbol{\theta}_0)T_j) +   \exp( - \lambda_i(\boldsymbol{\theta}_0)T_{j-1})\right)^{\beta-1}
\end{align*}
%\begin{align*}
%	 (\boldsymbol{z}_j-\boldsymbol{z}_{j-1})\pi_j(\boldsymbol{\theta})^{\beta-1} =& 
%	\left( g_T(T_j )\begin{pmatrix}
%		\frac{T_j}{\theta_0}\\
%		T_jx_i - a_{i-1}^\ast
%	\end{pmatrix} - g_T(T_{j-1} )\begin{pmatrix}
%	\frac{T_{j-1}}{\theta_0}\\
%	T_{j-1}x_i - a_{i-1}^\ast
%\end{pmatrix} 
%\right)(G_i(T_j)-G_i(T_{j-1}))^{\beta-1}\\
%=& 
%\bigg[ \theta_0 \exp(\theta_1x_i - \theta_0 \exp(\theta_1x_i)T_j)\begin{pmatrix}
%	\frac{T_j}{\theta_0}\\
%	T_jx_i - a_{i-1}^\ast
%\end{pmatrix} \\ 
% &-  \theta_0 \exp(\theta_1x_i - \theta_0 \exp(\theta_1x_i)T_{j-1})\begin{pmatrix}
%	\frac{T_{j-1}}{\theta_0}\\
%	T_{j-1}x_i - a_{i-1}^\ast
%\end{pmatrix} 
%\bigg]\\
%& \left( - \exp( - \theta_0 \exp(\theta_1x_i)T_j) +   \exp( - \theta_0 \exp(\theta_1x_i)T_{j-1})\right)^{\beta-1}
%\end{align*}
with $T_j = t_j+a_{i-1}-\tau_{i-1}.$ All the terms depending on times before $t_j$ are bounded, and the values $\lambda_i(\boldsymbol{\theta}_0), a_{i-1}, a_{i-1}^\ast$ and $\tau_{i-1}$ are positive constants. Then, taking limits as $T_j\rightarrow \infty,$ we get 
$$
\lim_{t_j \rightarrow \infty}(\boldsymbol{z}_j-\boldsymbol{z}_{j-1})\pi_j(\boldsymbol{\theta}_0)^{\beta-1} = \begin{cases}
	+ \infty & \text{if } \beta = 0,\\
	< \infty & \text{if } \beta > 0.\\
\end{cases}
$$
%A similar reasoning gives the convergence of the consecutive terms, $l=j+1,..L+1,$ yielding to the same result. 
Hence, the IF of the MDPDEs, for positive values of $\beta,$ is bounded when  any inspection time gets increased, whereas the IF of the MLE is unbounded for this class of leverage points.

Similarly, let us consider a stress level $x_i$ and let $x_i \rightarrow \infty.$ We take $t_j$ such that $t_j = \tau_i,$ the time of stress change for the $i$-th stress level. 
Again, as the stress levels are ordered, we can consider that the devices at subsequent steps are subjected to the same stress $x_i$. Then, we need to stablish the boundedness of all terms from $j$ onwards.
The lifetime rates are not constant since they depend on the stress level. Therefore, taking limits on (\ref{eq:loglinear}), we have
$$\lim_{ x_i \rightarrow \infty} \lambda_i(\boldsymbol{\theta}) = \begin{cases}
	0 & \text{if } \theta_1 \leq 0,\\
	\infty & \text{if } \theta_1 > 0.\\
\end{cases}$$
The limiting behaviour of the IF of $\theta_0$ and $\theta_1$ may be different, since the first only depends on the stress level at $g_T(T_j),$ whereas the IF of $\theta_1$ includes a term in $g_T(T_j)T_jx_i.$
Therefore, for $\theta_1 >0,$
\begin{align*}
	&\lim_{ x_i \rightarrow \infty} \left( g_i(T_j )\begin{pmatrix}
		\frac{T_j}{\theta_0}\\
		T_jx_i + a_{i-1}^\ast
	\end{pmatrix} - g_i(T_{j-1} )\begin{pmatrix}
		\frac{T_{j-1}}{\theta_0}\\
		T_{j-1}x_i + a_{i-1}^\ast
	\end{pmatrix}
	\right)(G_i(T_j)-G_i(T_{j-1}))^{\beta-1} \\
	=& 
	\lim_{ x_i \rightarrow \infty} \bigg( \lambda_i(\boldsymbol{\theta}_0)\exp(- \lambda_i(\boldsymbol{\theta}_0) T_j)\begin{pmatrix}
		\frac{T_j}{\theta_0}\\
		T_jx_i + a_{i-1}^\ast
	\end{pmatrix}  
	-  \lambda_i(\boldsymbol{\theta}_0) \exp(- \lambda_i(\boldsymbol{\theta}_0)T_{j-1})\begin{pmatrix}
		\frac{T_{j-1}}{\theta_0}\\
		T_{j-1}x_i + a_{i-1}^\ast
	\end{pmatrix} 
	\bigg)\\
	& \hspace{1.2cm} \times \left( - \exp( - \lambda_i(\boldsymbol{\theta}_0)T_j) +   \exp( - \lambda_i(\boldsymbol{\theta}_0)T_{j-1})\right)^{\beta-1}\\
	=& \begin{cases}
		- \infty & \text{if } \beta = 0,\\
		< \infty & \text{if } \beta > 0.\\
	\end{cases}
\end{align*}
For $\theta_1<0,$  we must deal with the IF of $\theta_0$ and the IF of $\theta_1$ separately. For the IF of the first parameter $\theta_0,$ we have
 \begin{equation*}
	\lim_{ x_i \rightarrow \infty} \left( g_i(T_j )
	\frac{T_j}{\theta_0} - g_i(T_{j-1} )
	\frac{T_{j-1}}{\theta_0}
	\right)(G_i(T_j)-G_i(T_{j-1}))^{\beta-1} < \infty \hspace{0.3cm}\forall \beta \geq 0,
\end{equation*}
whereas taking limits in the IF of the second parameter $\theta_1,$ we obtain
\begin{equation*}
	\lim_{ x_i \rightarrow \infty} \left( g_i(T_j)(T_jx_i + a_{i-1}^\ast)
	 - g_i(T_{j-1} )(T_{j-1}x_i + a_{i-1}^\ast)
	\right)(G_i(T_j)-G_i(T_{j-1}))^{\beta-1} = \begin{cases}
		+ \infty & \text{if } \beta = 0,\\
		< \infty & \text{if } \beta > 0.\\
	\end{cases} 
\end{equation*}
Thus, the IF of the proposed MDPDE is bounded for all $\beta > 0$ regardless of the sign of the true parameter value $\theta_1.$ In contrast, the IF of the MLE of the parameter $\theta_0$ is unbounded for positive true parameter values of $\theta_1$ and so is the IF of the MLE of the parameter $\theta_1$. This means that the proposed estimators are also robust for all type of outliers, whereas the MLE lacks robustness against this ``bad'' leverage points.
%of any MDPDE for the first parameter $\theta_0$ with $\beta\geq0$ is bounded,
%$ (\boldsymbol{z}_j-\boldsymbol{z}_{j-1})\pi_j(\boldsymbol{\theta})^{\beta-1}$ tends to zero for all whereas the IF 
\end{remark}

\section{Point estimation and confidence intervals of reliability and mean lifetime} \label{sec:interval}

One may be interested in studying the reliability of non-destructive one-shot devices or in estimating its expected lifetime. Technically, the reliability of a device is the probability that it will perform its intended function, under operating condition, for a specified period of time. Then, the reliability can be measured as the probability of survival until a pre-specified time under normal operating conditions.
Following the notation in Section \ref{sec:modelformulation}, the reliability or survival function of the lifetime $T$ of a device is given by 
\begin{equation*}
	R_T(t) = 1- G_T(t) = \begin{cases} R_1(t) = e^{-\lambda_1 t}, & 0<t< \tau_1,\\
		R_2(t + a_1-\tau_1)= e^{-\lambda_2(t + a_1 -\tau_1)}, & \tau_1 \leq t < \tau_2, \\
		\vdots \\
		R_K(t+a_{k-1}-\tau_{k-1}) = e^{-\lambda_k(t + a_{k-1} -\tau_{k-1})}, & \tau_{k-1} \leq t < \infty, \\
	\end{cases}
\end{equation*}
where $a_{i-1}$ is as defined in (\ref{eq:ai}) and $R_i(t), i = 1,...k,$ is the reliability function at the $i-th$ stress level, which depends in turn on the model parameters $\boldsymbol{\theta} = (\theta_0, \theta_1)$. Therefore, an estimated reliability at a certain time can be obtained from the above formula. 
%Note that the reliability of a device will depend on the stress level at which it is subjected. 
For cumulative exposure model with exponential lifetime distributions, the reliability of the device at the inspection interval $[\tau_i, \tau_{i+1}]$ (assuming that the stress level will not be increased) corresponds to the reliability function of a translated exponential  distribution with parameter $\lambda_i, i = 1,...,k$.
If the device is subjected to a constant stress level $x_i$, then its reliability at time $t$ can be computed as the reliability function $R_i(t).$ 
Let us denote $x_0$ for the stress level at normal operating conditions. Then, the reliability of the device is given by, for a fixed time $t,$ 
\begin{equation}\label{eq:survival}
	R_0(t) = \exp\left(-\lambda_0 t\right) = \exp\left(-\theta_0\exp(\theta_1x_0) t\right).
\end{equation}

Also, for planning purposes, one may need to estimate the time at which more than a certain percentage of devices are expected to fail under normal operating conditions. Mathematically, those times are the distribution quantiles, computed as the inverse distribution (or reliability) function,
\begin{equation}\label{eq:quantiles}
	Q_{1-\alpha}  =	R_0^{-1}(1-\alpha) = G_0^{-1}(\alpha) = -\frac{\log(1-\alpha)}{\lambda_0},
\end{equation}
%$$
%	\begin{cases}
%	\frac{-\log(1-\alpha)}{\lambda_k} - a_{k-1} +\tau_{k-1} & 0 \leq 1-\alpha < e^{-\lambda_k(\tau_{k-1}+a_{k-1}-\tau_{k-1})}  \\
%	\vdots \\ 
%	\frac{-\log(1-\alpha)}{\lambda_2}-a_1 +\tau_1 & e^{-\lambda_2(\tau_2+a_1-\tau_1)} \leq 1-\alpha <  e^{-\lambda_1\tau_1}  \\
%	\frac{-\log(1-\alpha)}{\lambda_1}  & e^{-\lambda_1\tau_1}< 1-\alpha<1 \\	
%\end{cases}
%$$
with $1-\alpha$ being the proportion of surviving units.
Further, the mean lifetime of a device under an exponential lifetime distribution with parameter $\lambda$ is $ \mathbb{E}[T] = 1/\lambda.$ From (\ref{eq:loglinear}), the mean lifetime of the device depends on the stress level through a log-linear relationship as
\begin{equation*}
	 \frac{1}{\lambda_i} = \frac{1}{\theta_0} \exp\left(-\theta_1 x_i\right), i = 1,...,k.
\end{equation*}
Hence, under normal operating conditions, the expected lifetime of the device is simply
\begin{equation}\label{eq:loglinearmean}
	\operatorname{E}_T = \mathbb{E}[T]= \frac{1}{\lambda_0} = \frac{1}{\theta_0} \exp\left(-\theta_1 x_0\right).
\end{equation}

Given the MDPDEs of the model parameters, its is straightforward to obtain point estimate of the reliability at a mission time, estimate quantiles and mean lifetime of the devices  under normal operating conditions by substituting the estimated parameters in (\ref{eq:survival})-(\ref{eq:loglinearmean}), yielding the estimators $\widehat{R}_0^\beta(t),$ $\widehat{Q}_{1-\alpha}^\beta$ and $\widehat{\operatorname{E}}_T^\beta,$ respectively.
Additional interest may be on confidence intervals (CI) for such quantities.
We first present the asymptotic distribution of the reliability, quantiles and mean lifetime estimators based on the MDPDEs, $\widehat{\boldsymbol{\theta}}^\beta$, under normal operating conditions. 
These results can be obtained readily from the asymptotic distribution of the MDPDE by employing the Delta method.

\begin{theorem} \label{thm:asymptoticreliability}
	Let $\boldsymbol{\theta}_0$ be the true value of the parameter $\boldsymbol{\theta}.$ Let $\widehat{\boldsymbol{\theta}}^\beta$ be the MDPDE, with tuning parameter $\beta.$
	Then, the asymptotic distribution of the estimated reliability
	at a mission time $t,$ under normal operating conditions,   based on the MDPDE $\widehat{\boldsymbol{\theta}}^\beta$, $\widehat{R}_0^\beta(t),$ is given by
	$$\sqrt{N}(\widehat{R}_0^\beta(t)- R_0^\beta(t)) \xrightarrow[N \rightarrow \infty]{L} \mathcal{N}\left(\boldsymbol{0}, \sigma(R_0^\beta(t))^2 \right) ,$$
	with
	$$\sigma(R_0^\beta(t))^2= \nabla h\left(\boldsymbol{\theta}_0\right)^T \boldsymbol{J}_\beta^{-1}(\boldsymbol{\theta}_0)\boldsymbol{K}_\beta(\boldsymbol{\theta}_0)\boldsymbol{J}_\beta^{-1}(\boldsymbol{\theta}_0)\nabla h\left(\boldsymbol{\theta}_0\right),$$
	where the matrices 
  $\boldsymbol{J}_\beta(\boldsymbol{\theta}_0)$ and $\boldsymbol{K}_\beta(\boldsymbol{\theta}_0)$ are as defined in (\ref{eq:JK}) and 
  $\nabla h\left(\boldsymbol{\theta}\right)^T = \left(- R_0(t)  \frac{\lambda_0 t}{\theta_0}, - R_0(t)\lambda_0 t x_0\right)$ is the gradient of the function $h(\boldsymbol{\theta}) = \exp(-\theta_0\exp(\theta_1x_0)t).$
\end{theorem}

%\begin{proof}
%	Since the MDPDE $\widehat{\boldsymbol{\theta}}^\beta$ has an asymptotic normal distribution,
%	$$ \sqrt{N}\left(\boldsymbol{\widehat{\theta}}^{\beta} - \boldsymbol{\theta}_0\right) \rightarrow \mathcal{N}\left(\boldsymbol{0}, \boldsymbol{J}_\beta^{-1}(\boldsymbol{\theta}_0)\boldsymbol{K}_\beta(\boldsymbol{\theta}_0)\boldsymbol{J}_\beta^{-1}(\boldsymbol{\theta}_0)\right),$$
%	%with matrices $\boldsymbol{J}_\beta(\boldsymbol{\theta}_0)$ and $\boldsymbol{K}_\beta(\boldsymbol{\theta}_0)$ defined in (\ref{eq:JK}).
%	 we can apply the Delta-method to obtain the asymptotic distribution of $h(\widehat{\theta_0},\widehat{\theta_1}) = \exp(-\widehat{\theta}_0\exp(\widehat{\theta}_1x_0)t) = \widehat{R}_0^\beta(t).$
%	 
%\end{proof}

\begin{theorem}\label{thm:asymptoticquantiles}
	Under the same assumptions as in Result \ref{thm:asymptoticreliability}, the asymptotic distribution of the estimated $(1-\alpha)$- quantile,
	under normal operating conditions, based on the MDPDE $\widehat{\boldsymbol{\theta}}^\beta$, $\widehat{Q}_{1-\alpha}^\beta,$ is given by
	$$\sqrt{N}(\widehat{Q}_{1-\alpha}^\beta- Q_{1-\alpha}) \xrightarrow[N \rightarrow \infty]{L} \mathcal{N}\left(\boldsymbol{0}, \sigma(Q_{1-\alpha})^2 \right) ,$$
	with
	$$\sigma(Q_{1-\alpha})^2 = \nabla h_1\left(\boldsymbol{\theta}_0\right)^T \boldsymbol{J}_\beta^{-1}(\boldsymbol{\theta}_0)\boldsymbol{K}_\beta(\boldsymbol{\theta}_0)\boldsymbol{J}_\beta^{-1}(\boldsymbol{\theta}_0)\nabla h_1\left(\boldsymbol{\theta}_0\right) $$
	where the matrices 
	$\boldsymbol{J}_\beta(\boldsymbol{\theta}_0)$ and $\boldsymbol{K}_\beta(\boldsymbol{\theta}_0)$ are as defined in (\ref{eq:JK}) and 
	$$\nabla h_1\left(\boldsymbol{\theta}\right)^T = \left( \frac{\log(1-\alpha)}{\theta_0^2}\exp(-\theta_1x_0), \frac{\log(1-\alpha)x_0}{\theta_0}\exp(-\theta_1x_0) \right)$$ is the gradient of the function $h_1(\boldsymbol{\theta}) = \frac{1}{\theta_0}\exp(-\theta_1x_0).$
\end{theorem}

%\begin{proof}
%	The proof is similar than in Theorem (\ref{thm:asymptoticreliability}).
%\end{proof}

\begin{theorem}\label{thm:asymptoticmean}
	Under the same assumptions as in Result \ref{thm:asymptoticreliability}, the asymptotic distribution of the estimated mean lifetime,
	 under normal operating conditions, based on the MDPDE $\widehat{\boldsymbol{\theta}}^\beta$, $\widehat{\operatorname{E}_T}^\beta,$ is given by
	$$\sqrt{N}(\widehat{\operatorname{E}}_T^\beta- \operatorname{E}_T) \xrightarrow[N \rightarrow \infty]{L} \mathcal{N}\left(\boldsymbol{0}, \sigma(\operatorname{E}_T)^2 \right) ,$$
	with
	$$\sigma(\operatorname{E}_T)^2 = \nabla h_2\left(\boldsymbol{\theta}_0\right)^T \boldsymbol{J}_\beta^{-1}(\boldsymbol{\theta}_0)\boldsymbol{K}_\beta(\boldsymbol{\theta}_0)\boldsymbol{J}_\beta^{-1}(\boldsymbol{\theta}_0)\nabla h_2\left(\boldsymbol{\theta}_0\right)$$
	where the matrices 
	$\boldsymbol{J}_\beta(\boldsymbol{\theta}_0)$ and $\boldsymbol{K}_\beta(\boldsymbol{\theta}_0)$ are as defined in (\ref{eq:JK}) and 
	$\nabla h_2\left(\boldsymbol{\theta}\right)^T = \left( \frac{-1}{\theta_0^2}\exp(-\theta_1x_0), \frac{-x_0}{\theta_0}\exp(-\theta_1x_0) \right)$ is the gradient of the function $h_2(\boldsymbol{\theta}) = \frac{1}{\theta_0}\exp(-\theta_1x_0).$
\end{theorem}
%
%\begin{proof}
%	The proof is similar than in Theorem (\ref{thm:asymptoticreliability}).
%\end{proof}

As $\widehat{\boldsymbol{\theta}}^\beta$ are consistent estimators, from the above results, we can easily obtain approximate two-sided $100(1 - \alpha)\%$ CI for the reliability, $(1-\alpha)$-quantile and mean lifetime, under normal operating conditions, to be
$$\widehat{R}_0^\beta(t) \pm  z_{\alpha/2}\frac{\sigma(\widehat{R}_0^\beta(t))}{\sqrt{N}}, \hspace{0.3cm}
\widehat{Q}_{1-\alpha}^\beta \pm  z_{\alpha/2}\frac{\sigma(\widehat{Q}_{1-\alpha}^\beta)}{\sqrt{N}}
\hspace{0.3cm} \text{ and } \hspace{0.3cm} \widehat{\operatorname{E}}_{T}^\beta \pm  z_{\alpha/2}\frac{\sigma(\widehat{\operatorname{E}}_T^\beta)}{\sqrt{N}}$$
where $\sigma(\widehat{R}_0^\beta(t)),$ $\sigma(\widehat{Q}_{1-\alpha}^\beta)$ and $\sigma(\widehat{\operatorname{E}}_T^\beta)$ are as defined in Results \ref{thm:asymptoticreliability}-\ref{thm:asymptoticmean}, respectively.
%CI of such measures relies on the CI of the MDPDEs, that can be obtained by different methods. 
%In particular, we analyze approximate and  bootstrap methods for constructing CI.

%Let $[l_i,u_i]$ be the $100(1 - \alpha)\%$ CI of $\theta_i$ based on the MDPDE $\widehat{\theta}_i^\beta, i = 0,1,$ obtained by any of the previous methods. Without loss of generality, we can assume that $l_i > 0, i = 0,1.$ Then, the CI of the reliability of the devices at a certain time $t$ is given by

%Similarly, the $100(1 - \alpha)\%$ CI of the mean lifetime, derived from the MPDPE $\widehat{\boldsymbol{\theta}}^\beta,$ at a certain stress level $x_i$ is given by $$\frac{1}{u_0}\exp(-u_1) \leq \frac{1}{\lambda_i} \leq \frac{1}{l_0}\exp(-l_1).$$
%With large sample sizes, both CI reach similar results

%Reliability: Reliability is the probability that a product will operate or a service will be provided properly for a specified period of time (design life) under the design operating conditions (such as temperature, load, volt…) without failure

%$$R(t) = P(T > t)$$
%The hazard function is defined as the limit of the failure rate as Δt approaches zero

The above asymptotic confidence intervals are based on the asymptotic properties of the estimators and so they may be satisfactory only for large sample sizes. In small samples, we may have to truncate  the confidence intervals as the mean lifetime and quantiles must be positive and the reliability should be between 0 and 1. In this regard, Viveros and Balakrishnan (1993) employed a logit transformation of the estimated reliability to obtain more accurate CIs based on the MLE. The transformed reliability is defined as 
\begin{equation}\label{eq:logitreliability}
	\phi = \phi(R_0(t)) = \operatorname{logit}(R_0(t))= \log\left(\frac{R_0(t)}{1-R_0(t)}\right),
\end{equation}
where $\phi \in \mathbb{R}.$ Thus, the range for this transformed reliability would not require truncation. The logit transformation is a natural choice when dealing with parameters that represent probabilities, since it results in $\mathbb{R}$.
Estimated values of the transformed reliabilities based on the MDPDEs, $\widehat{\phi}^\beta,$ can be easily obtained by substituting the corresponding estimated reliabilities $\widehat{R}_0^\beta(t)$ in (\ref{eq:logitreliability}), and their asymptotic distribution and CIs can be derived by using  Delta method. Inverting such a CI, after some algebra, we obtain the asymptotic CI for the reliability as
$$  \left[ \frac{\widehat{R}_0^\beta(t)}{\widehat{R}_0^\beta(t)+(1-\widehat{R}_0^\beta(t))S}, \frac{\widehat{R}_0^\beta(t)}{\widehat{R}_0^\beta(t)+(1-\widehat{R}_0^\beta(t))/S}\right],
$$
with $ S= \exp\left( \frac{z_{\alpha/2}}{\sqrt{N}}\frac{\sigma(\widehat{R}_0^\beta(t))}{\widehat{R}_0^\beta(t)(1-\widehat{R}_0^\beta(t))}\right)$ and $\sigma(R_0(t))$ as defined in Result \ref{thm:asymptoticreliability}.

A similar idea can be applied for the quantiles and mean lifetimes in (\ref{eq:quantiles}) and (\ref{eq:loglinearmean}). As both quantities must be positive, the logarithm is a natural choice for transforming them to $\mathbb{R}$. Transformed quantiles and mean lifetimes are then
\begin{equation}
	\phi_1 = \log\left(Q_{1-\alpha}\right) \text{ and } 	\phi_2 = \log\left(\operatorname{E}_T\right).
\end{equation}
Again, using Delta method for deriving the asymptotic distributions, and then inverting the logarithmic transformations, we obtain CIs for $Q_{1-\alpha}$ and $\operatorname{E}_T$ as
$$\left[\widehat{Q}_{1-\alpha}^\beta\exp\left( -\frac{z_{\alpha/2}}{\sqrt{N}}\frac{\sigma(\widehat{Q}_{1-\alpha}^\beta)}{\widehat{Q}_{1-\alpha}^\beta}\right), \widehat{Q}_{1-\alpha}^\beta \exp\left( \frac{z_{\alpha/2}}{\sqrt{N}}\frac{\sigma(\widehat{Q}_{1-\alpha}^\beta)}{\widehat{Q}_{1-\alpha}^\beta}\right) \right]$$
%with $S_1 = \exp\left( \frac{z_{\alpha/2}}{N}\frac{\sigma(\widehat{\operatorname{E}}_T^\beta)}{\widehat{\operatorname{E}}_T^\beta}\right)$
and 
$$\left[\widehat{\operatorname{E}}_T^\beta\exp\left( -\frac{z_{\alpha/2}}{\sqrt{N}}\frac{\sigma(\widehat{\operatorname{E}}_T^\beta)}{\widehat{\operatorname{E}}_T^\beta}\right), \widehat{\operatorname{E}}_T^\beta \exp\left( \frac{z_{\alpha/2}}{\sqrt{N}}\frac{\sigma(\widehat{\operatorname{E}}_T^\beta)}{\widehat{\operatorname{E}}_T^\beta}\right) \right],$$
respectively, with $\sigma(\operatorname{E}_T)$ and $\sigma(Q_{1-\alpha})$ as defined in Results \ref{thm:asymptoticquantiles} and \ref{thm:asymptoticmean}.

\section{Robust tests of hypotheses} \label{sec:robusttest}

In this section, we consider linear hypothesis tests on the model parameter $\boldsymbol{\theta},$ of the form
\begin{equation}\label{eq:null}\operatorname{H}_0 : \boldsymbol{m}^T\boldsymbol{\theta} = d,
\end{equation}
where $\boldsymbol{m} =(m_0,m_1)^T \in \mathbb{R}^2$. In particular,  the linear hypothesis with $\boldsymbol{m}^T = (0,1)$ and $d=0$ would test if the stress level affects the lifetime of the one-shot devices or not.
We present here the a testing procedure based on the MDPDE and then we study it robustness and asymptotic behaviour.

%\subsection{Z-type test}
Specifically, we define the Z-type statistics based on the MDPDE and then study theoretically its asymptotic distribution under the null and contiguous hypotheses and robustness properties
\begin{definition} The Z-type statistic based on the MDPDE $\widehat{\boldsymbol{\theta}}^\beta,$ for testing null hypothesis (\ref{eq:null}), is given by
	\begin{equation}\label{eq:Ztypest}
		Z_{N}(\widehat{\boldsymbol{\theta}}^\beta) = \sqrt{N}\left(\boldsymbol{m}^T\boldsymbol{J}_\beta^{-1}(\widehat{\boldsymbol{\theta}}^\beta)\boldsymbol{K}_\beta(\widehat{\boldsymbol{\theta}}^\beta)\boldsymbol{J}_\beta^{-1}(\widehat{\boldsymbol{\theta}}^\beta)\boldsymbol{m}\right)^{-1/2}\left(\boldsymbol{m}^T\widehat{\boldsymbol{\theta}}^\beta - d\right).
	\end{equation}
\end{definition}
 The asymptotic distribution of this statistic is given in the following result
\begin{theorem}\label{thm:asymptoticstat}
	The asymptotic distribution of the Z-type statistic (\ref{eq:Ztypest}), under the null hypothesis (\ref{eq:null}), is a standard normal distribution.
\end{theorem}

Based on Result \ref{thm:asymptoticstat}, for any $\beta \geq 0$ and $\boldsymbol{m} \in \mathbb{R}^2,$ the critical region with significance level $\alpha$ for the hypothesis test with linear null hypothesis in (\ref{eq:null}), with level $\alpha$ is given by
\begin{equation}\label{eq:criticalregionZtest}
	\mathcal{R}_{\alpha} = \{(n_1,...,n_{L+1})  \text{ s.t. } | Z_{N}(\widehat{\boldsymbol{\theta}}^\beta)| > z_{\alpha/2}\},
\end{equation}
where $z_{\alpha/2}$ denotes the upper $\alpha/2$-quantile of the standard normal distribution.

\begin{remark}
	We can generalize the null hypothesis in (\ref{eq:null}) to
	\begin{equation*}
		\operatorname{H}_0 : \boldsymbol{M}^T\boldsymbol{\theta} = \boldsymbol{d}
	\end{equation*}
	with $\boldsymbol{M}$ being a $r\times2$ ($r\leq2$) matrix and $\boldsymbol{d}$ being a $r-$dimensional vector. Then, we can define the corresponding test statistic as
	\begin{equation}
		Z^\ast_{N}(\widehat{\boldsymbol{\theta}}^\beta) = N\left(\boldsymbol{M}^T\widehat{\boldsymbol{\theta}}^\beta - \boldsymbol{d}\right)^T \left(\boldsymbol{M}^T\boldsymbol{J}_\beta^{-1}(\widehat{\boldsymbol{\theta}}^\beta)\boldsymbol{K}_\beta(\widehat{\boldsymbol{\theta}}^\beta)\boldsymbol{J}_\beta^{-1}\boldsymbol{M}\right)^{-1}\left(\boldsymbol{M}^T\widehat{\boldsymbol{\theta}}^\beta - \boldsymbol{d}\right).
	\end{equation}
	Note that the matrix $\boldsymbol{M}^T\boldsymbol{J}_\beta^{-1}(\widehat{\boldsymbol{\theta}}^\beta)\boldsymbol{K}_\beta(\widehat{\boldsymbol{\theta}}^\beta)\boldsymbol{J}_\beta^{-1}(\widehat{\boldsymbol{\theta}}^\beta)\boldsymbol{M}$ is symmetric, and so the statistic is well defined.
	It is not difficult to stablish that, under the generalized null hypothesis above,
	$$\sqrt{N}\left(\boldsymbol{M}^T\boldsymbol{J}_\beta^{-1}(\widehat{\boldsymbol{\theta}}^\beta)\boldsymbol{K}_\beta(\widehat{\boldsymbol{\theta}}^\beta)\boldsymbol{J}_\beta^{-1}(\widehat{\boldsymbol{\theta}}^\beta)\boldsymbol{m}\right)^{-1/2}\left(\boldsymbol{M}^T\widehat{\boldsymbol{\theta}}^\beta - \boldsymbol{d}\right) \xrightarrow[L\rightarrow \infty]{L} \mathcal{N}(\boldsymbol{0}, \boldsymbol{I}),$$  
	and therefore, 
	$$
	 Z^\ast_{N}(\widehat{\boldsymbol{\theta}}^\beta) \xrightarrow[L\rightarrow \infty]{L} \chi_r^2.$$
	Consequently, a critical region corresponding to  the generalized null hypothesis is
	$$ \mathcal{R}_{\alpha} = \{(n_1,...,n_{L+1})  \text{ s.t. }  Z^\ast_{N}(\widehat{\boldsymbol{\theta}}^\beta) > \chi_{r,\alpha}^2\},$$
	where $\chi_{r,\alpha}^2$ denotes the upper $\alpha$-quantile of a chi-square distribution with $r$ degrees of freedom.
\end{remark}

The robustness of the proposed Z-type test statistic can be established by its IF. The IF of a testing procedure at a contamination point $\boldsymbol{n}$ is defined as the Gateaux derivative of the functional, defining the test statistic at the contamination direction given by $\Delta_{\boldsymbol{n}}.$ 
In the present context, the functional associated with the proposed Z-type test statistic, $Z_{N}(\widehat{\boldsymbol{\theta}}^\beta),$ under the null hypothesis is given by
\begin{equation*}
	Z_{N}(\boldsymbol{T}_\beta(G)) = \sqrt{\frac{N}{\boldsymbol{m}^T\boldsymbol{J}_\beta^{-1}(\boldsymbol{\theta}_0)\boldsymbol{K}_\beta(\boldsymbol{\theta}_0)\boldsymbol{J}_\beta^{-1}(\boldsymbol{\theta}_0)\boldsymbol{m}}}\left(\boldsymbol{m}^T\boldsymbol{T}_\beta(G) - d\right).
\end{equation*}
Therefore, the IF of the proposed Z-type test statistic can be easily derived from the IF of the MDPDE, as
\begin{align*}
	\text{IF}\left(\boldsymbol{n}, Z_{N}, G\right) &= \frac{\partial Z_{N}(\boldsymbol{T}_\beta(G_\varepsilon))}{\partial \varepsilon}\bigg|_{\varepsilon = 0}\\
	& = \sqrt{\frac{N}{\boldsymbol{m}^T\boldsymbol{J}_\beta^{-1}(\boldsymbol{\theta}_0)\boldsymbol{K}_\beta(\boldsymbol{\theta}_0)\boldsymbol{J}_\beta^{-1}(\boldsymbol{\theta}_0)\boldsymbol{m}}}\boldsymbol{m}^T \frac{\partial \boldsymbol{T}_\beta(G_\varepsilon)}{\partial \varepsilon}\bigg|_{\varepsilon = 0}\\
	& = \sqrt{\frac{N}{\boldsymbol{m}^T\boldsymbol{J}_\beta^{-1}(\boldsymbol{\theta}_0)\boldsymbol{K}_\beta(\boldsymbol{\theta}_0)\boldsymbol{J}_\beta^{-1}(\boldsymbol{\theta}_0)\boldsymbol{m}}}\boldsymbol{m}^T \text{IF}\left(\boldsymbol{n}, \boldsymbol{T}_\beta, G\right).
\end{align*}
The boundedness of the IF of the Z-type test statistic at a contamination point $\boldsymbol{n}$ and the true distribution $F_{\boldsymbol{\theta}_0}$ can be discussed by the boundedness of the IF of the corresponding MDPDE, and thus, robust estimators results in robust test statistics.

%\section{Alternatives hypothesis}
On the other hand, we can obtain the asymptotic distribution of the Z-type test in (\ref{eq:Ztypest}) at a contiguous alternative hypothesis.
Let $\boldsymbol{\theta}_L \in \Theta \setminus \Theta_0$ be an alternative and take $\boldsymbol{\theta}_0$  as the  closest element to the boundary of $\Theta_0$ in the sense of Euclidean distance.
We consider contiguous alternative hypothesis  of the form
\begin{equation}\label{eq:contiguous}
	\operatorname{H}_{1,L}: \boldsymbol{\theta} = \boldsymbol{\theta}_L,
\end{equation}
with $ \boldsymbol{\theta}_L = \boldsymbol{\theta}_0 + \frac{1}{\sqrt{N}}\ell,$ for a fixed vector $\ell \in \mathbb{R}^2$. 
% close to the null hypothesis.
Note that, defining $\ell^{\ast} = \boldsymbol{m}^T\ell,$ we have
$$\boldsymbol{m}^T\boldsymbol{\theta}_L - d = \boldsymbol{m}^T(\boldsymbol{\theta}_L - \boldsymbol{\theta}_0) = \boldsymbol{m}^T\frac{\ell}{\sqrt{N}}$$
so that the contiguous hypothesis in (\ref{eq:contiguous}) can be equivalently stated by  the condition  $g(\boldsymbol{\theta}_L) = \frac{1}{\sqrt{N}}\ell^\ast.$ 
%Let consider the 
\begin{theorem}
	The asymptotic distribution of the Z-type statistic in (\ref{eq:Ztypest}),  under the contiguous hypothesis (\ref{eq:contiguous}),
	is a normal distribution, with mean $(\boldsymbol{m}^T\boldsymbol{J}_\beta^{-1}(\widehat{\boldsymbol{\theta}}^\beta)\boldsymbol{K}_\beta(\widehat{\boldsymbol{\theta}}^\beta)\boldsymbol{J}_\beta^{-1}(\widehat{\boldsymbol{\theta}}^\beta)\boldsymbol{m})^{-1/2}\boldsymbol{m}^T\ell$ and unit variance.
		%$$\left(\boldsymbol{m}^T\boldsymbol{J}_\beta^{-1}(\boldsymbol{\theta}_0)\boldsymbol{K}_\beta(\boldsymbol{\theta}_0)\boldsymbol{J}_\beta^{-1}(\boldsymbol{\theta}_0)\boldsymbol{m}\right)^{-1/2}\left(\sqrt{N}(\boldsymbol{m}^T\widehat{\boldsymbol{\theta}}^\beta - d) - \boldsymbol{m}^T \ell\right) \xrightarrow[L\rightarrow\infty]{} \mathcal{N}\left( \boldsymbol{0} , \boldsymbol{I}_{r\times r}\right)$$
\end{theorem}

From the above result, we can obtain an approximation for the power function of the test statistic in (\ref{eq:null}) at the contiguous hypothesis in (\ref{eq:contiguous}), as
%Under these contiguous hypothesis, the Z-type test statistic are asymptotically biased.
\begin{align*}
	\beta_N\left(\boldsymbol{\theta}_L\right) &=  \mathbb{P}\left(|Z_{N}(\widehat{\boldsymbol{\theta}}^\beta)| > z_{\alpha/2} | \boldsymbol{\theta} = \boldsymbol{\theta}_L \right)\\
	&\approx 2\left[1-\Phi\left(z_{\alpha/2}-\sqrt{\frac{N}{\boldsymbol{m}^T\boldsymbol{J}_\beta^{-1}(\widehat{\boldsymbol{\theta}}^\beta)\boldsymbol{K}_\beta(\widehat{\boldsymbol{\theta}}^\beta)\boldsymbol{J}_\beta^{-1}(\widehat{\boldsymbol{\theta}}^\beta)\boldsymbol{m}}}\boldsymbol{m}^T\ell \right)\right]
\end{align*}

It is clear that $\lim_{N\rightarrow \infty} \beta_N(\boldsymbol{\theta}_L)=1$ and so the Z-type statistic is consistent in the sense of Fraser (1957). More generally, the following result provides an asymptotic approximation to the power function.

\begin{theorem}
	Let $\boldsymbol{\theta}^\ast \in \Theta$ be the true value of the parameter $\boldsymbol{\theta}$ with $\boldsymbol{m}^T\boldsymbol{\theta}^\ast \neq d.$ Then, the approximate power function of the test statistic
	in (\ref{eq:null}) is given by
	$$\beta_N\left(\boldsymbol{\theta}^\ast\right) \approx 2\left[1-\Phi\left(1-\sqrt{N}\left(\boldsymbol{m}^T\boldsymbol{J}_\beta^{-1}(\widehat{\boldsymbol{\theta}}^\beta)\boldsymbol{K}_\beta(\widehat{\boldsymbol{\theta}}^\beta)\boldsymbol{J}_\beta^{-1}(\widehat{\boldsymbol{\theta}}^\beta)\boldsymbol{m}\right)^{-1/2} \left(\boldsymbol{m}^T\boldsymbol{\theta}^\ast -d \right)\right)\right],$$
	where $\Phi(\cdot)$ denotes the standard normal distribution function.
\end{theorem}

\section{Simulation study}\label{sec:simstudy}

In this section, we examine the behaviour of the proposed robust MDPDEs, Z-type tests and Rao-type tests.
% MDPDE, $\widehat{\boldsymbol{\theta}}^\beta,$ and the Z-type and Rao-type test statistics, $Z_{N}(\widehat{\boldsymbol{\theta}}^\beta)$ and $R_{\beta, N}(\widetilde{\boldsymbol{\theta}}^\beta)$
 for the SSALT model with exponential lifetime  distribution under different contamination scenarios.

For multinomial sampling, we must consider ``outlying cells'' instead of  ``outlying devices'', see Balakrishnan et al. (2019a). Then, to introduce contamination in our context, we should increase (or decrease) the probability of failure in (\ref{eq:th.prob}) for (at least) one interval (i.e., one cell). So, the probability of failure is switched in such contaminated cells as
\begin{equation} \label{eq:contaminatedprob}
	\tilde{\pi}_j(\boldsymbol{\theta}) =  G_{\boldsymbol{\theta}}(\textit{IT}_j) - G_{\boldsymbol{\tilde{\theta}}}(\textit{IT}_{j-1})
\end{equation}
for some $j=2,...,L$, where $\boldsymbol{\tilde{\theta}} = (\widetilde{\theta}_0,\widetilde{\theta}_1)$ is a contaminated parameter with $\tilde{\theta}_0 \leq \theta_0$  and $\tilde{\theta}_1 \leq \theta_1.$ It is important to point out that, after the contamination of the probability of failure in a cell, the probability vector of the multinomial model must be normalized to add up to 1. 
%explain that model with great outliers tarnish the estimation

\subsection{Minimum density power divergence estimators} \label{sec:simstudyMDPPE}

Let us consider a 2-step stress ALT experiment with $L=11$ inspection times and a total of $N = 180$
devices under test. At the beginning of the experiment, all the devices are subjected to a stress level $x_1=35$ until the first time of stress change $\tau_1 = 25.$ Then, the surviving units are subjected to an increased stress level, $x_2=45,$ till the end of the experiment at $\tau_2=70.$ During the experiment, inspection is performed at a grid of inspection times containing the times of stress change, $\text{IT}=(10,15,20,25,30,35,40,45,50,60,70).$
We set the true value of the true parameter $\boldsymbol{\theta}_0=(0.003,0.03),$ and then generate data from the corresponding
multinomial model described in Section \ref{sec:modelformulation}, with exponential lifetimes. Moreover, we contaminate the data by increasing the probability of failure in the third interval as mentioned in (\ref{eq:contaminatedprob}).

In order to evaluate the performance of the proposed estimators, we calculate the root mean square error (RMSE) of the MDPDE for different values of $\beta \in \{0,0.2,0.4,0.6,0.8,1\},$ including the MLE for $\beta = 0.$ Further,  to asses the efficiency loss of an estimator with respect to the MLE, we 
define a measure $\rho(\widehat{\boldsymbol{\theta}}^\beta),$ quantifying the relative RMSE of an estimator with respect to the RMSE of the MLE, as
$$\rho(\widehat{\boldsymbol{\theta}}^\beta) = \frac{||\widehat{\boldsymbol{\theta}}^\beta-\boldsymbol{\theta}_0||_2}{||\widehat{\boldsymbol{\theta}}^0-\boldsymbol{\theta}_0||_2}-1.$$
Then, $\rho(\widehat{\boldsymbol{\theta}}^\beta)$ measures the efficiency loss of an estimator with respect to the MLE. Clearly, when $\rho(\widehat{\boldsymbol{\theta}}^\beta) <0,$ the MDPDE is more accurate than the MLE, and evidently $\rho(\widehat{\boldsymbol{\theta}}^0) =1.$

In addition, we test different scenarios of contamination. In the first scenario, we generate an outlying cell in the third interval by decreasing the value of the first parameter, $\theta_0,$ and in the second scenario we perform similarly, but decreasing the second parameter $\theta_1$. In both cases, the lifetime rate $\lambda(\widetilde{\boldsymbol{\theta}}),$ is decreased; the smaller is the contamination parameter, the greater is the contamination. 
%use the ratio between the RMSE of the corresponding estimator and the RMSE of the MLE. More concretely, we define a measure of efficiency loss as 

\begin{figure}[H]
	\begin{subfigure}{0.5\textwidth}
		\includegraphics[height=8cm, width=8cm]{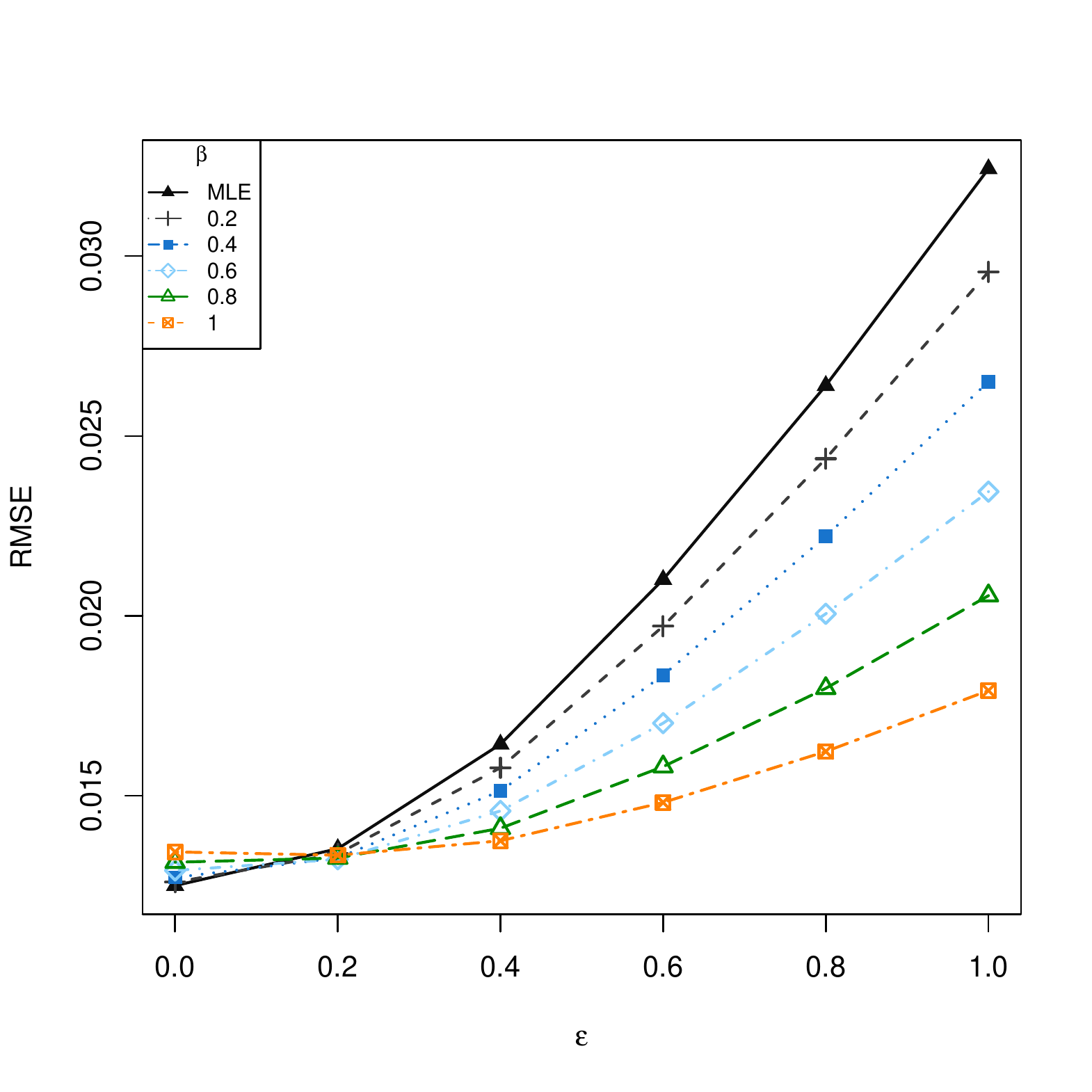}
		\subcaption{$\theta_0$-contaminated cell}
	\end{subfigure}
	\begin{subfigure}{0.5\textwidth}
		\includegraphics[height=8cm, width=8cm]{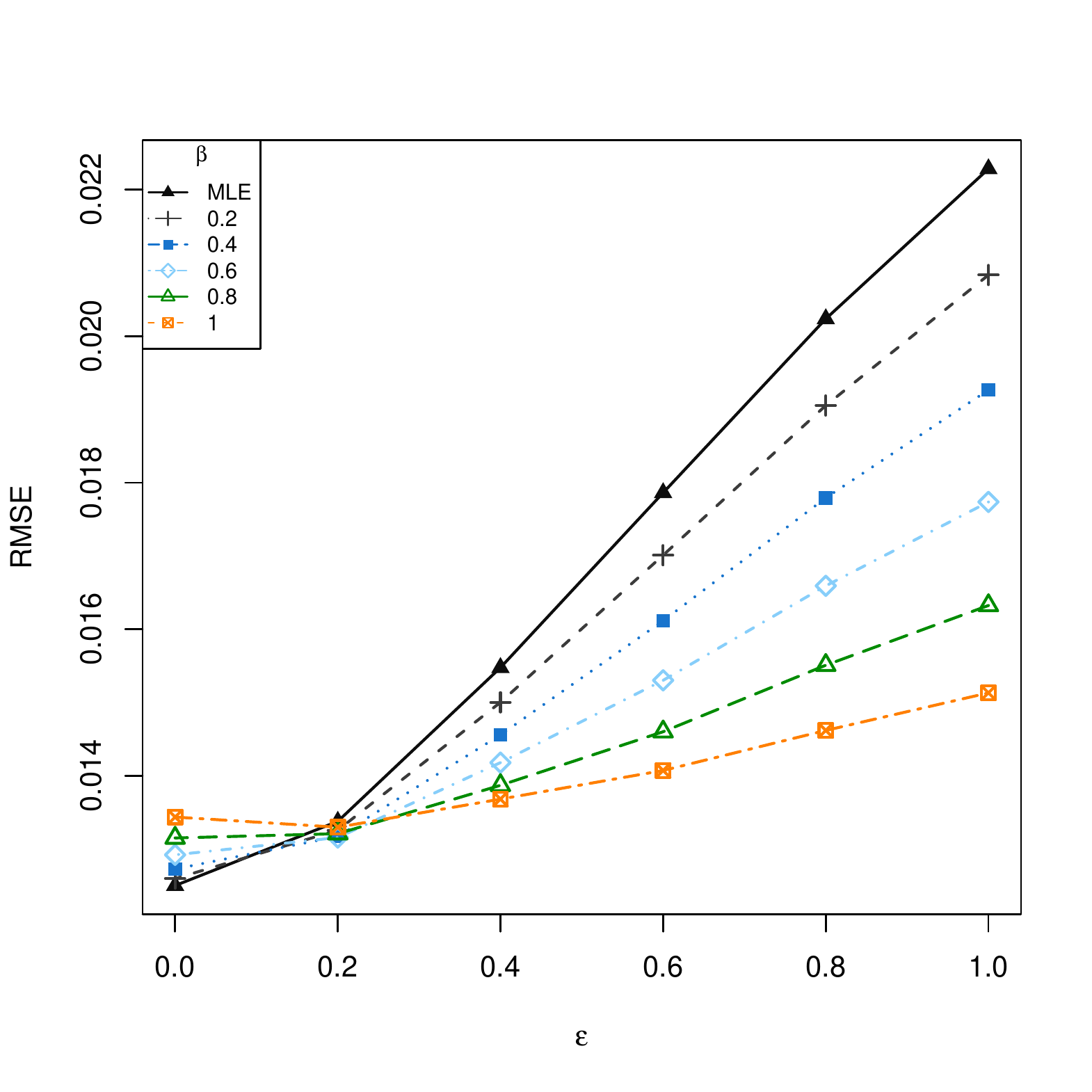}
		\subcaption{$\theta_1$-contaminated cell}
	\end{subfigure}
	\caption{RMSE of different estimators against data contamination in $R=1000$ replications}
	\label{fig:RMSE}
\end{figure}

\begin{figure}[H]
	\begin{subfigure}{0.5\textwidth}
		\includegraphics[height=8cm, width=8cm]{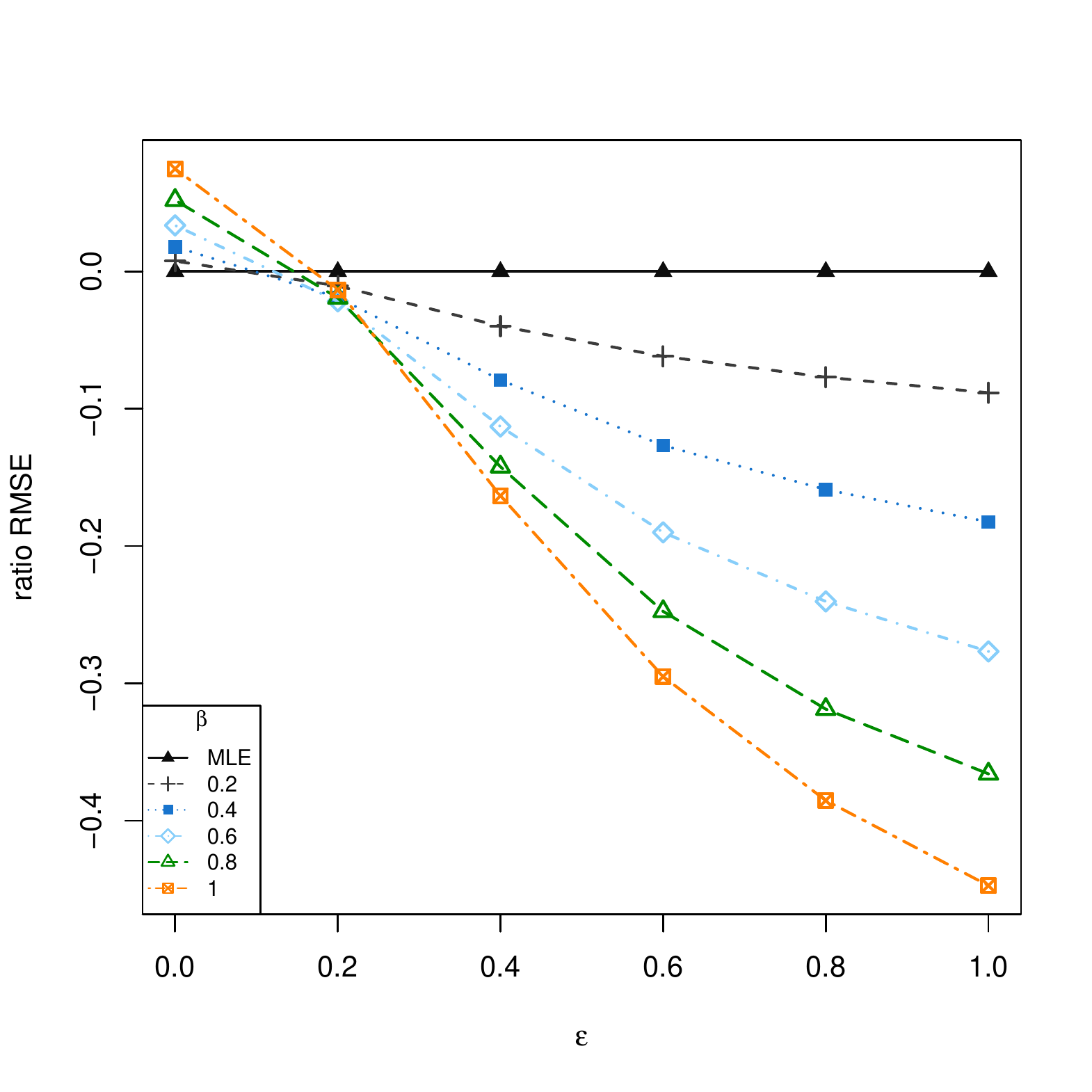}
		\subcaption{$\theta_0$-contaminated cell}
	\end{subfigure}
	\begin{subfigure}{0.5\textwidth}
		\includegraphics[height=8cm, width=8cm]{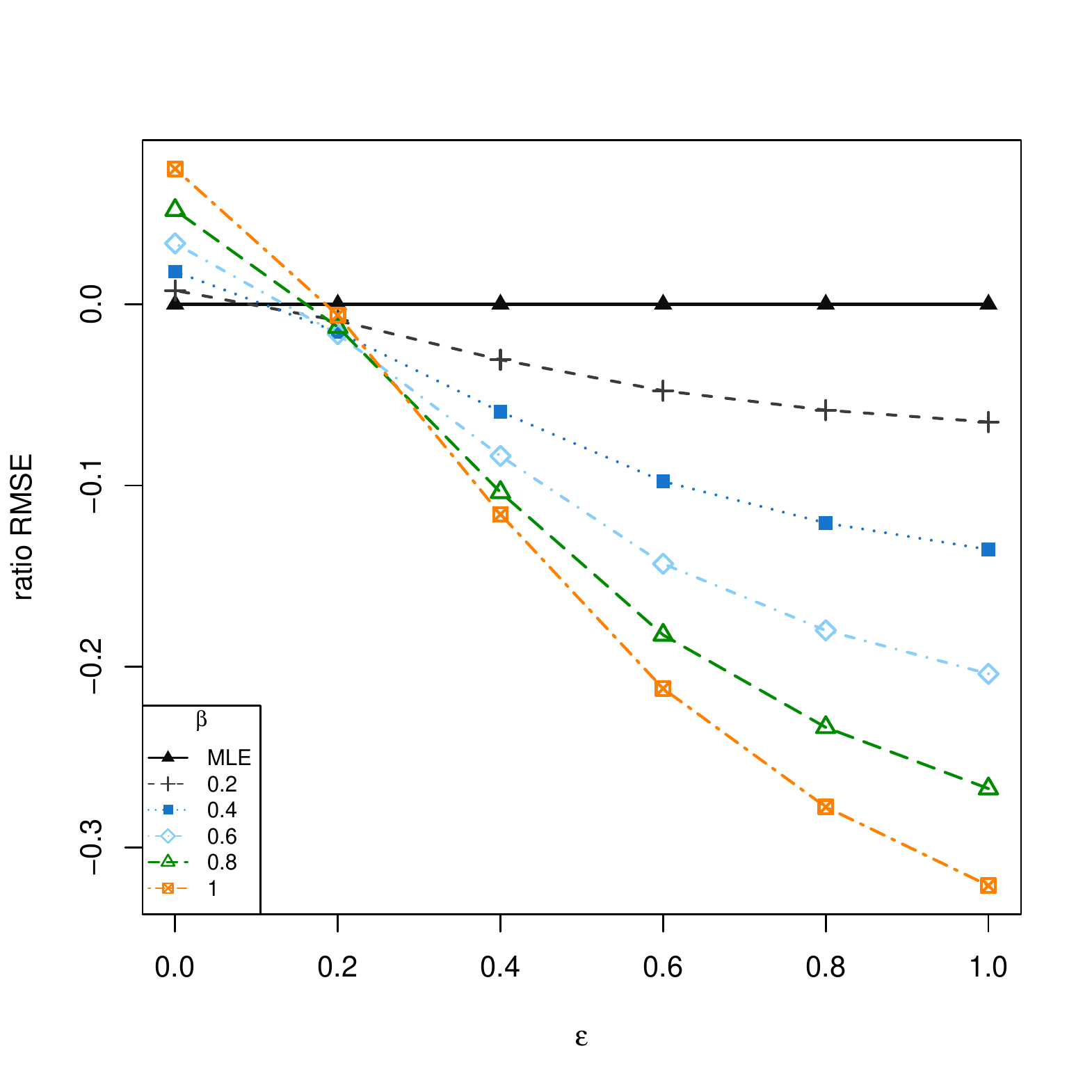}
		\subcaption{$\theta_1$-contaminated cell}
	\end{subfigure}
	\caption{Efficiency loss with respect to the MLE against data contamination in $R=1000$ replications}
	\label{fig:RMSEratio}
\end{figure}

Figures \ref{fig:RMSE} and \ref{fig:RMSEratio} show the RMSE and the RMSE ratio, $\rho,$ produced with different values of $\beta$ and the two contamination scenarios determined from $R=1000$ replications. In the left side plots, the contamination  is introduced by decreasing $\theta_0,$ yielding a contamination rate of $\varepsilon = 1- \frac{\widehat{\theta}_0}{\theta_0}$ while on the right side plots they are computed by decreasing the second parameter $\theta_1,$ and the corresponding contamination rate is then $\varepsilon = 1- \frac{\widehat{\theta}_1}{\theta_1}.$ 

These results show the advantage of the proposed MDPDE in terms of robustness. The larger is the parameter $\beta$, the more robust is the corresponding estimator. 
As expected, in the absence of contamination, the MLE is the most efficient estimator, even though all proposed MDPDEs perform competitively in this uncontaminated scenario. On the other hand, when a ``great'' outlier cell is generated, the efficiency loss of MLE with respect to the proposed MDPDE is seen to be quite pronounced. More specifically, under the contamination rates greater than $20\%,$ the MDPDEs outperform the MLE.

\subsection{Z-type tests} \label{sec:ztestsimulation}

\begin{figure}[H]
	\begin{subfigure}{0.5\textwidth}
		\includegraphics[height=8cm, width=8cm]{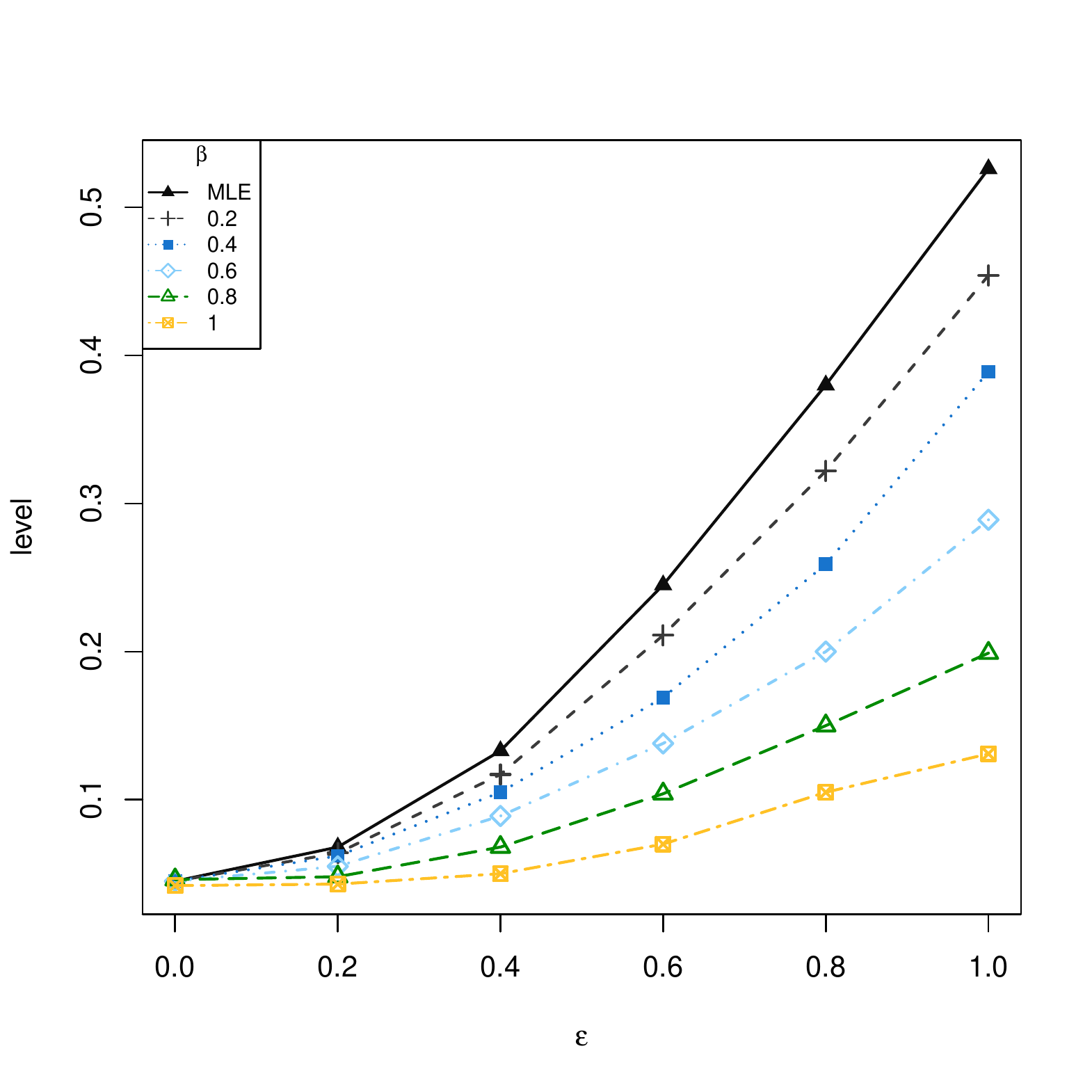}
		\subcaption{$\theta_0$-contaminated cell}
	\end{subfigure}
	\begin{subfigure}{0.5\textwidth}
		\includegraphics[height=8cm, width=8cm]{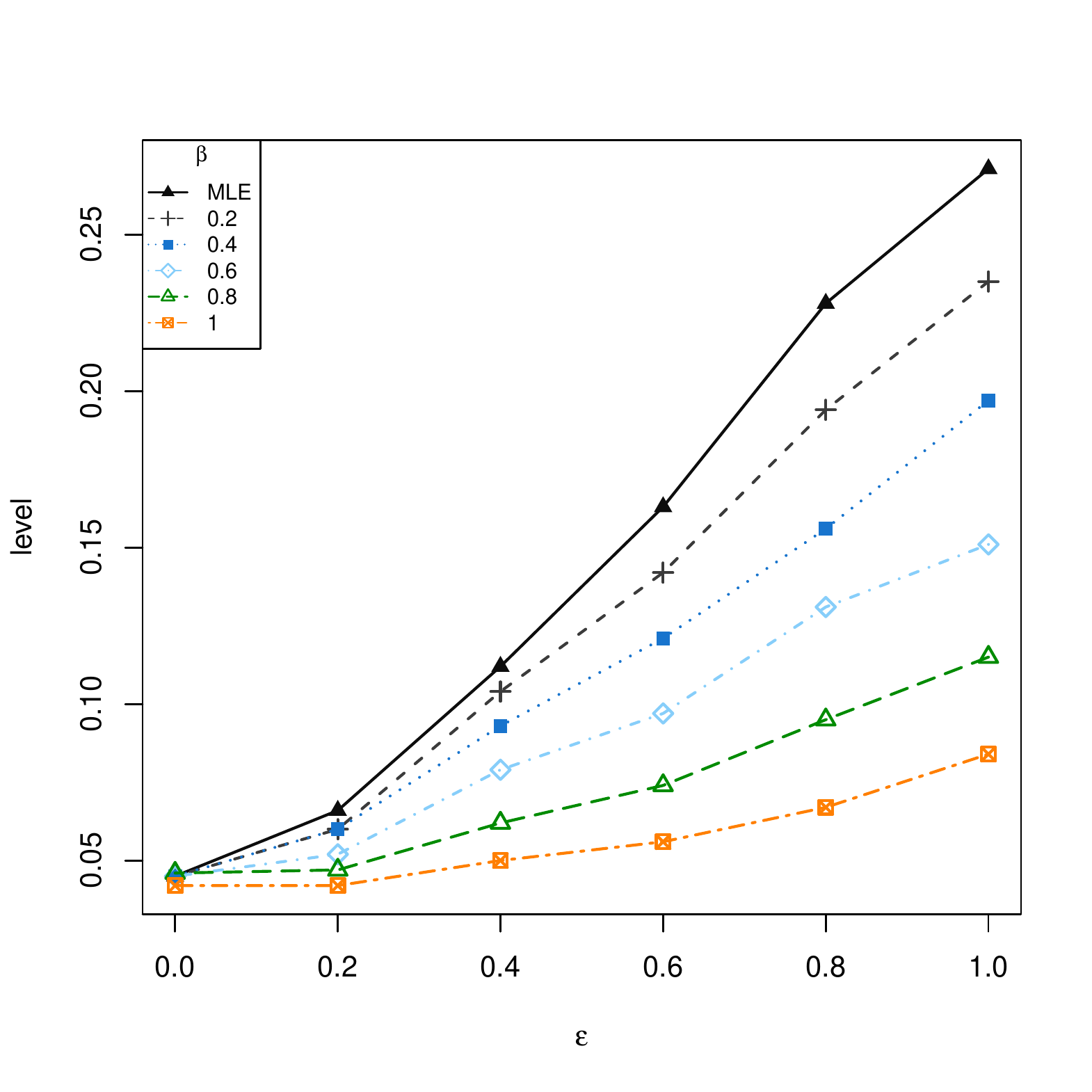}
		\subcaption{$\theta_1$-contaminated cell}
	\end{subfigure}
	\caption{Empirical significance level against contamination cell proportion in $R=1000$ replications}
	\label{fig:level}
\end{figure}

We empirically examine the performance of the Z-type test statistic based on the MDPDE. We adopt again the 2-step stress ALT experiment with $L=11$ inspection times and $N=180$ devices described in Section \ref{sec:simstudyMDPPE},
 and we consider testing the hypothesis
 %with null hypothesis
\begin{equation} \label{empiricaltest}
	\operatorname{H}_0: \theta_1=0.03 \hspace{0.5cm} \text{vs} \hspace{0.5cm} \operatorname{H}_1: \theta_1\neq0.03.
\end{equation} 
The true value of the parameter is set to be $\boldsymbol{\theta}_0 = (0.003,0.03)^T$ so as to fit the null hypothesis.
Then, the Z-type test statistic is defined using (\ref{eq:Ztypest}) with $\boldsymbol{m} = (0,1)^T$ and $d=0.03,$ and the critical region of the test is given by  (\ref{eq:criticalregionZtest}).
%$$ Z_{N}(\widehat{\boldsymbol{\theta}}^\beta) = \sqrt{N}\left((0,1)^T\boldsymbol{J}_\beta^{-1}(\widehat{\boldsymbol{\theta}}^\beta)\boldsymbol{K}_\beta(\widehat{\boldsymbol{\theta}}^\beta)\boldsymbol{J}_\beta^{-1}(\widehat{\boldsymbol{\theta}}^\beta)\boldsymbol{m}\right)^{-1/2}\left(\boldsymbol{m}^T\widehat{\boldsymbol{\theta}}^\beta - d\right) $$
Figure \ref{fig:level} shows the empirical  level of the test against cell contamination, $\varepsilon,$ with the two different contaminated scenarios considered in the last section, $\theta_0$-contaminated third cell (left) and $\theta_1$-contaminated third cell (right). The empirical level is computed as the proportion of rejected Z-type test statistic over $R=1000$ replications of the model under the null hypothesis for a significance level of $\alpha = 0.05$. The empirical level of the Z-type tests based on the MLE promptly grows  when the contamination rate gets increased, whereas the empirical level of the Z-type test based on the MDPDE, with large values of $\beta,$ remains low in heavily contaminated scenarios, highlighting its robustness property.

\begin{figure}[H]	\begin{subfigure}{0.5\textwidth}		\includegraphics[height=8cm, width=8cm]{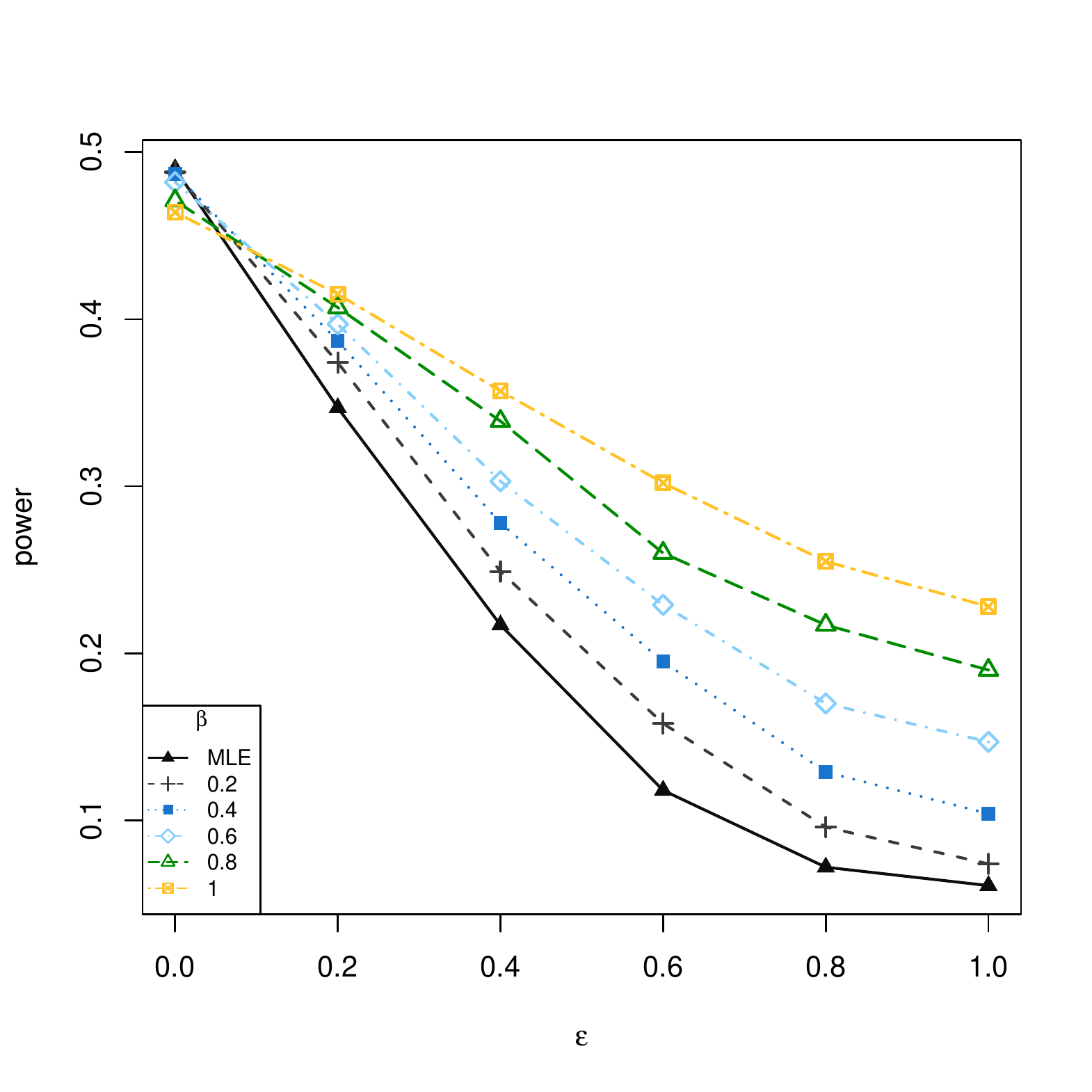}
		\subcaption{$\theta_0$-contaminated cell}
	\end{subfigure}
	\begin{subfigure}{0.5\textwidth}
		\includegraphics[height=8cm, width=8cm]{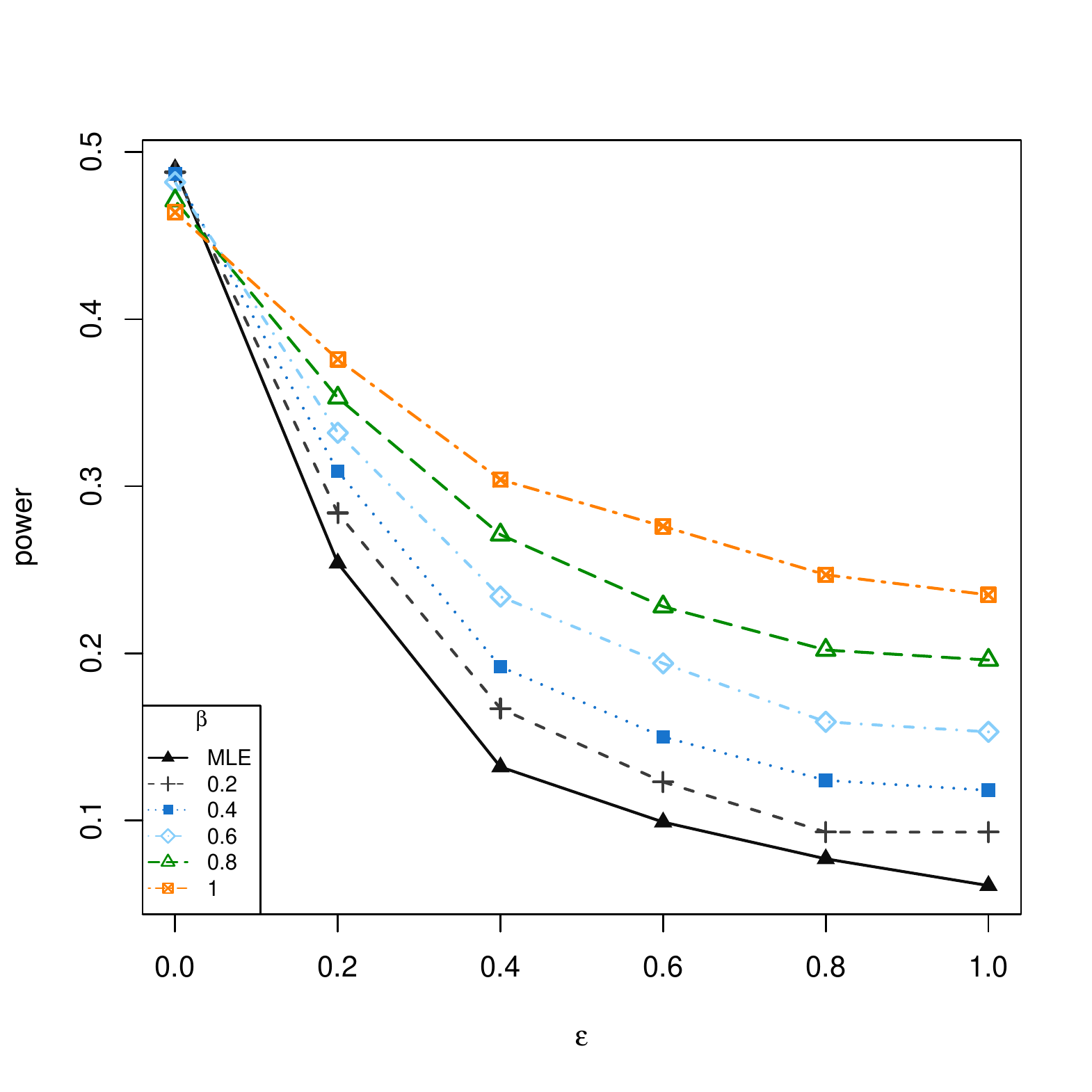}
		\subcaption{$\theta_1$-contaminated cell}
	\end{subfigure}
	\caption{Empirical power against contamination cell proportion in $R=1000$ replications}
	\label{fig:power}
\end{figure}

Next, we examine the empirical power of the linear hypothesis test by considering a different true parameter value, $\boldsymbol{\theta} = (0.003,0.6)^T,$ and the hypothesis test $\operatorname{H}_0: \theta_1=0.03$ against the alternative $\operatorname{H}_1: \theta_1\neq0.03.$ 
Now, the true value of the parameter does not satisfy the null hypothesis, and the Z-type test statistic has its the empirical power  under the two different contaminated scenarios as displayed in Figure \ref{fig:power}, with $R=1000$ replications. Again, the robustness of the Z-type test based on MDPDE becomes better when large values of $\beta$ are used to construct the test statistics.

\begin{figure}[H]
	\begin{subfigure}{0.5\textwidth}
		\includegraphics[height=7cm, width=8cm]{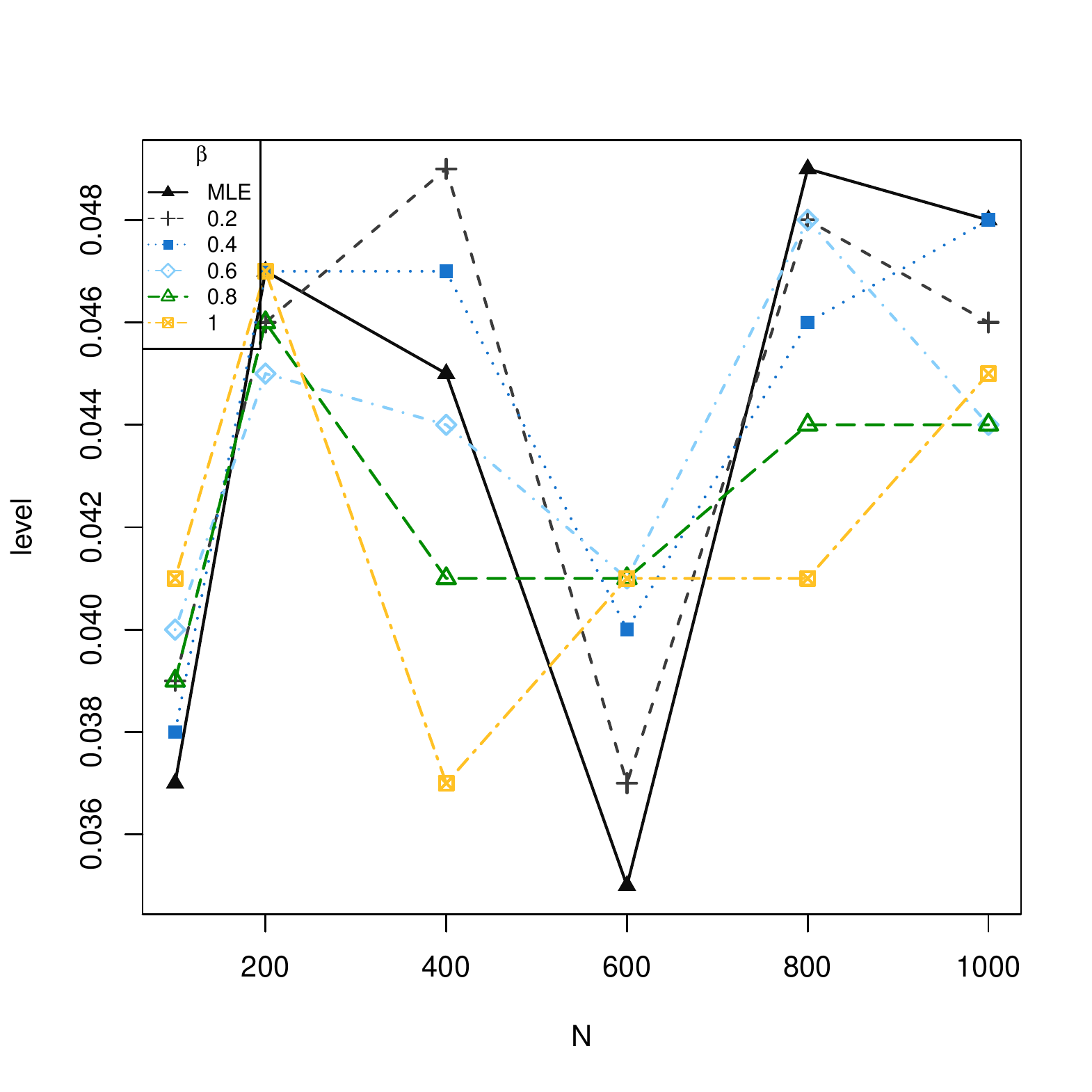}
		\subcaption{Absence of contamination}
	\end{subfigure}
	\begin{subfigure}{0.5\textwidth}
		\includegraphics[height=7cm, width=8cm]{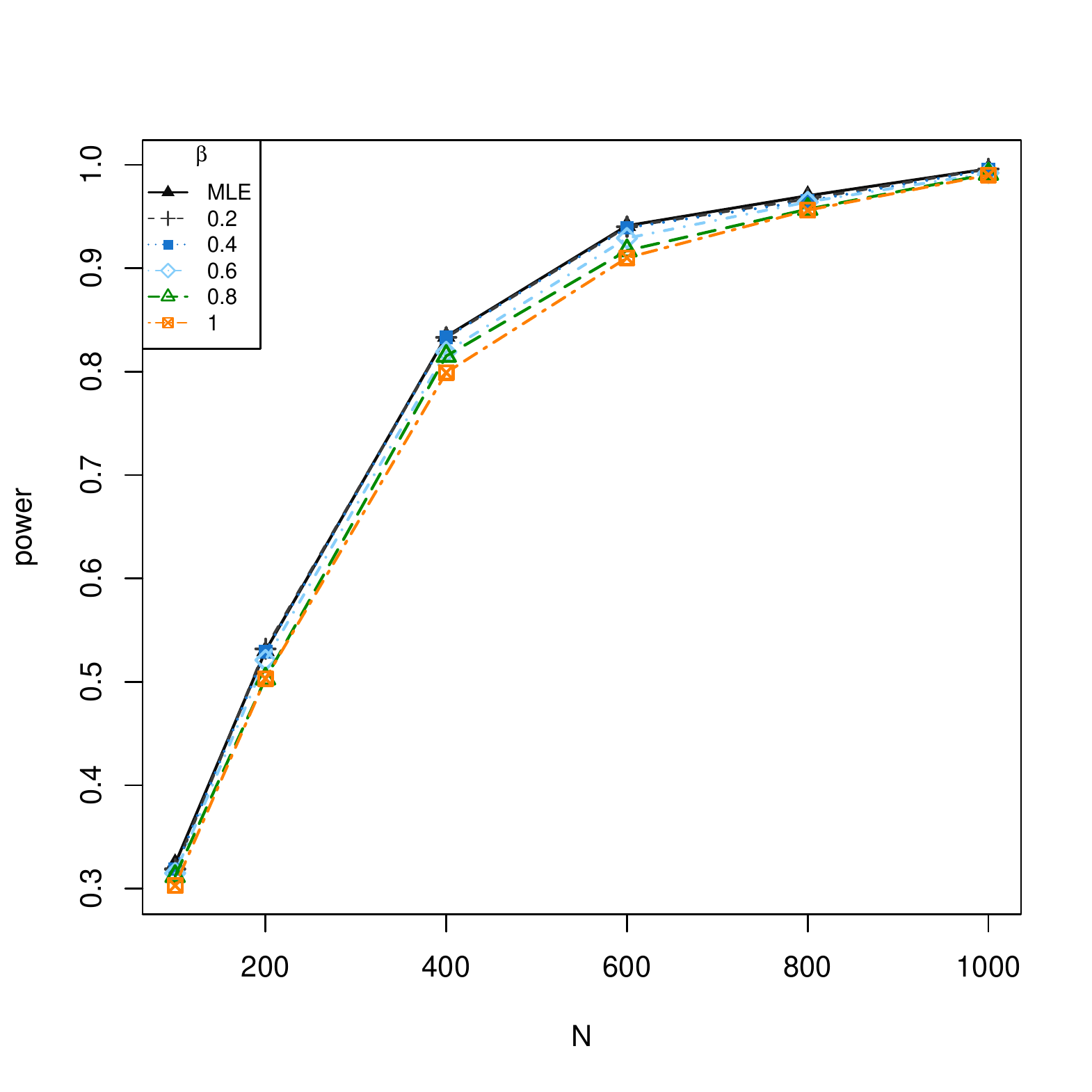}
		\subcaption{Absence of contamination}
	\end{subfigure}
	\begin{subfigure}{0.5\textwidth}
		\includegraphics[height=7cm, width=8cm]{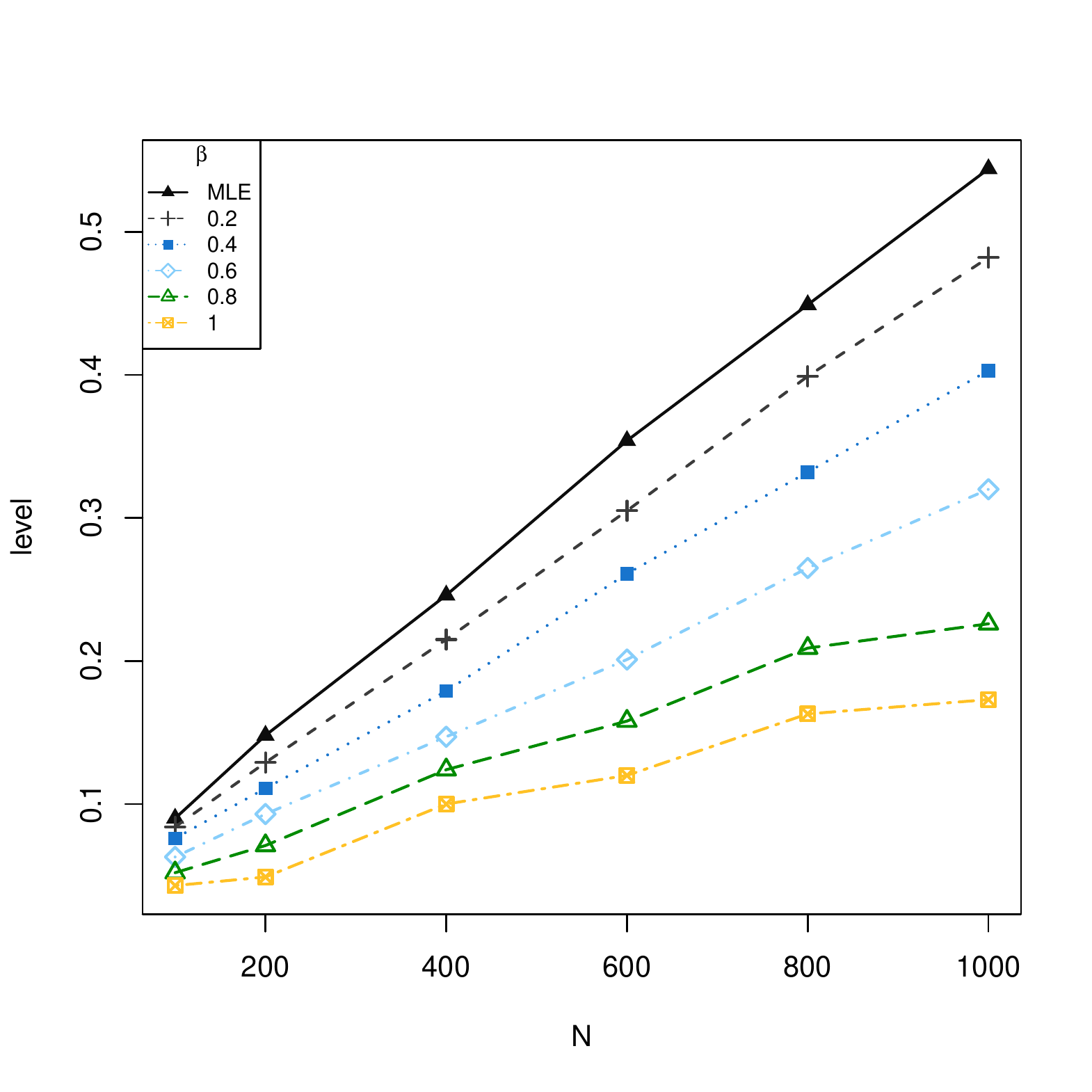}
		\subcaption{$\theta_0$-contaminated cell}
	\end{subfigure}
	\begin{subfigure}{0.5\textwidth}
	\includegraphics[height=7cm, width=8cm]{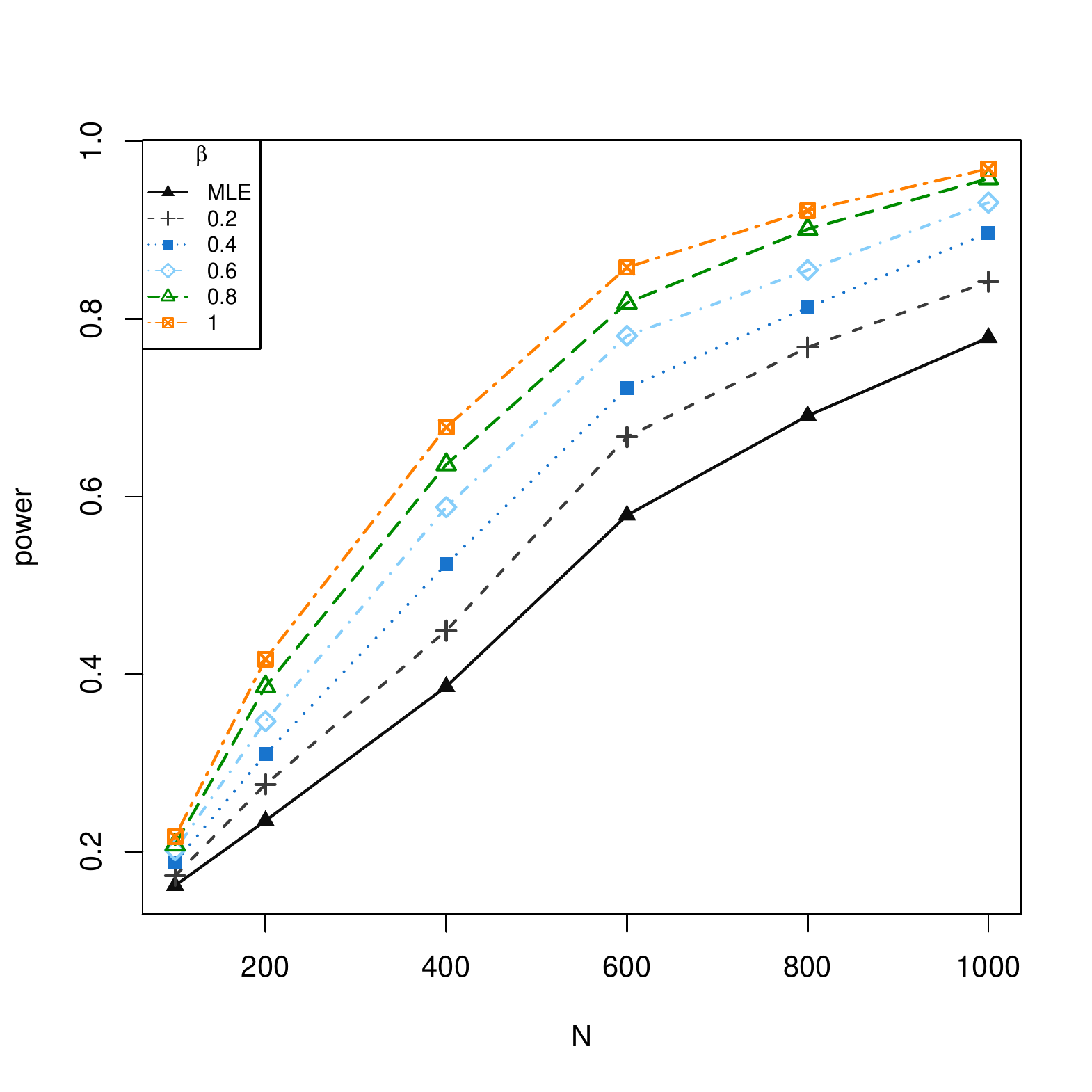}
	\subcaption{$\theta_0$-contaminated cell}
	\end{subfigure}
	\begin{subfigure}{0.5\textwidth}
		\includegraphics[height=7cm, width=8cm]{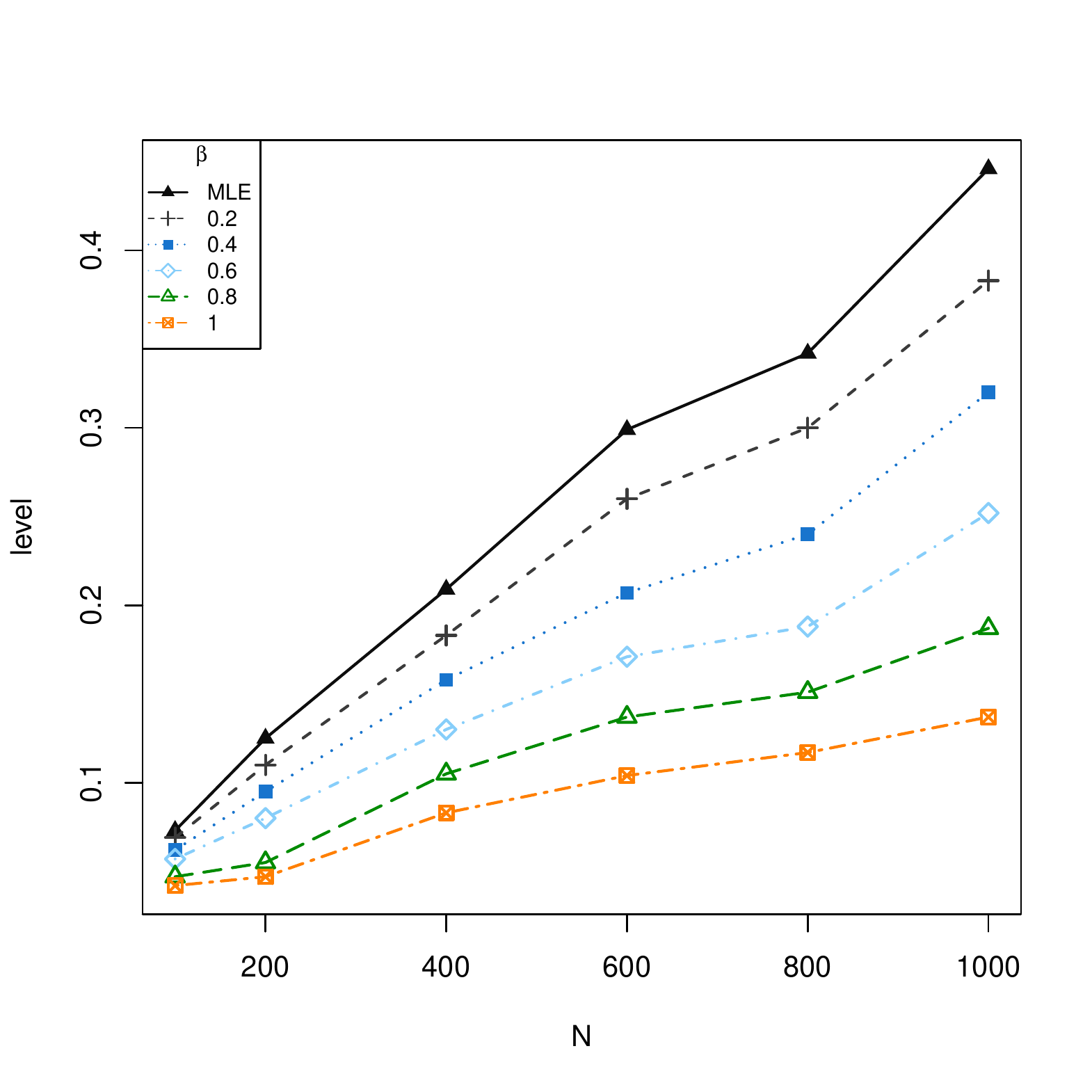}
		\subcaption{$\theta_1$-contaminated cell}
	\end{subfigure}
		\begin{subfigure}{0.5\textwidth}
		\includegraphics[height=7cm, width=8cm]{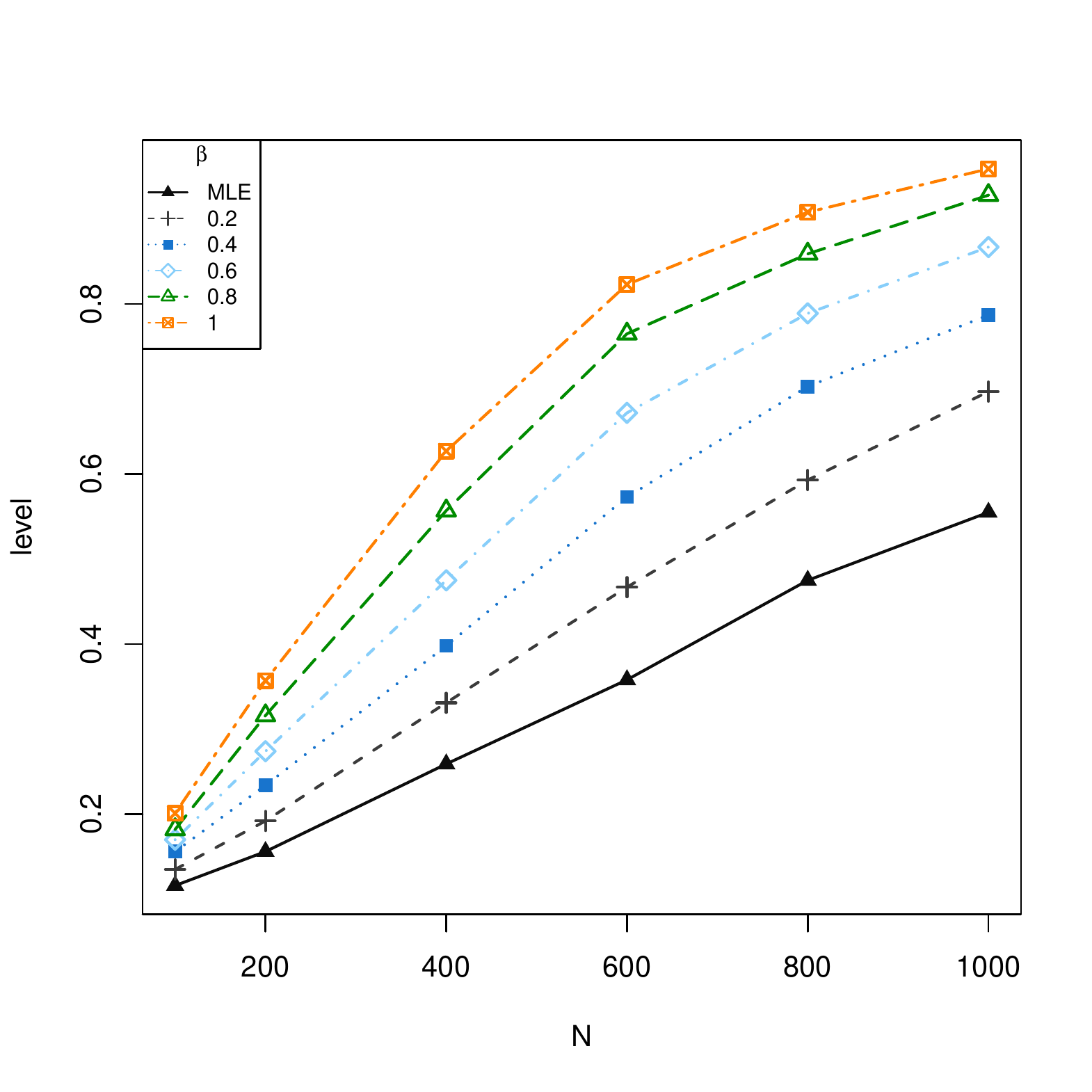}
		\subcaption{$\theta_1$-contaminated cell}
	\end{subfigure}
	\caption{Empirical level and power against sample size with different contaminated scenarios in $R=1000$ replications}
	\label{fig:levelandpowerN}
\end{figure}

We finally examine the performance of the Z-type test statistic against sample size with different contaminated scenarios:
in the absence of contamination (top), $\theta_0$-contaminated third cell with a $40\%$ reduction of the parameter (middle) and $\theta_1$-contaminated third cell with a $40\%$ reduction of the parameter (bottom). Figure \ref{fig:levelandpowerN} show the empirical level (left) and power (right) obtained with different sample sizes over $R=1000$ replications. In the absence of contamination (top), all Z-type tests based on MDPDEs with different values of $\beta$ perform similarly, even though the Z-type test based on the MLE is slightly better. On the other hand, when an outlying cell is introduced by decreasing any of the model parameters, the Z-type test based on MDPDE with larger values of $\beta$ outperform the test based on the MLE.
%We use the same set-up previously defined to compute the empirical level and power 

\subsection{Choice of the tuning parameter}\label{sec:choicebeta}

The tuning parameter $\beta$ of the DPD loss function controls the trade-off between efficiency and robustness of the resulting MDPPE. Following the discussions in the preceding sections, larger values of $\beta$ produce more robust but less efficient estimators.
Therefore,  the optimal value of $\beta$ will be good to determine. From our empirical results, a moderately large value of $\beta$ (over $\beta = 0.4$) is expected to provide robust estimators without a high loss of efficiency with respect to the MLE in the absence of contamination.
Determining this optimal value for the best compromise is, therefore, of great practical interest. Optimal values of $\beta$ will produce robust estimators without coming at the cost of a high efficiency loss. Then, a criterion measuring the efficiency loss  in favour of robustness gain should be adopted.

%Several criteria have been proposed in the literature.
 Warwick and Jones (2005) introduced an useful data-based procedure for the choice of the tuning parameter for the MDPDE. However, this method depends on the choice of a pilot estimator and Basak et al. (2021) improved the method by removing the dependency on an initial estimator. 
 The approach of Warwick and Jones (2005) minimizes the asymptotic MSE of the MDPDE given by
\begin{equation} \label{eq:choicebeta}
	\widehat{\operatorname{MSE}}\left(\beta\right) = \left(\widehat{\boldsymbol{\theta}}^\beta- \boldsymbol{\theta}_P\right)^T\left( \widehat{\boldsymbol{\theta}}^\beta- \boldsymbol{\theta}_P\right)
	+ \frac{1}{N} \operatorname{Tr}\left\{ \boldsymbol{J}_\beta(\widehat{\boldsymbol{\theta}}^\beta)^{-1} \boldsymbol{K}_\beta(\widehat{\boldsymbol{\theta}}^\beta) \boldsymbol{J}_\beta(\widehat{\boldsymbol{\theta}}^\beta)^{-1}\right\},
\end{equation}
 where $\boldsymbol{\theta}_P$ is a pilot estimator and $\operatorname{Tr}$ denote the trace of the matrix. Several proposals of this pilot estimator have been studied in the literature. However, the choice of the pilot significantly impact on the optimal tuning parameter, as it invariably draws the final estimator towards itself.
 To overcome this drawback, Basak et al. (2021) proposed an iterative algorithm that replaces in each step, the value of the pilot estimator by the estimator obtained with the optimal value of $\beta$ until the optimal choice of the tuning parameter (or equivalently, the pilot estimator) gets stabilized.
 The process should be initialized with a suitable robust pilot estimator, but the final choice of $\beta$ gets more pilot-independent.
 Basak et al. (2021) empirically showed that when the pilot estimators are within the MDPDE class, all robust pilots lead to the same iterated optimal choice, and moreover the performance of the algorithm improves even with pure data.
 These are all summarized in the following algorithm:\\
 
 \textbf{Algorithm} [Choice of the tuning parameter]
 \begin{enumerate}
 	 \item Fix the convergence rate $\varepsilon$ and choose an initial pilot estimator $\boldsymbol{\theta}_P$ from the MDPDE family;
 	 \item Update the optimal value of the tuning parameter, $\beta^\ast,$ using  the minimum asymptotic MSE in (\ref{eq:choicebeta});
 	 \item If the optimal estimate $\widehat{\boldsymbol{\theta}}^{\beta^\ast}$ differs from the pilot estimator by less than the convergence rate: Stop;\\
 	 Else, replace the pilot estimator by the optimal $\widehat{\boldsymbol{\theta}}^{\beta^\ast}$ and return to Step 2.
 \end{enumerate}

Let us consider the previous simulation set up with true parameter value $\boldsymbol{\theta} = (0.003,0.03)^T,$ but now let us choose the optimal value of $\beta$ according to the presented data-based procedure detailed in the above algorithm. We initialize the method with the MDPDE with different tuning parameters $\beta_p = 0, 0.5$ and $1$, yielding the pilot estimators $\widehat{\boldsymbol{\theta}}^{0}$, $\widehat{\boldsymbol{\theta}}^{0.5}$ and $\widehat{\boldsymbol{\theta}}^{1},$ respectively, and we fix the convergence rate to be $\varepsilon = 0.001.$  The minimization of (\ref{eq:choicebeta}) is carried out over a grid search in $[0,1]$ of size 100.

\begin{figure}[htb]
	\begin{subfigure}{0.5\textwidth}
		\includegraphics[height=8cm, width=8cm]{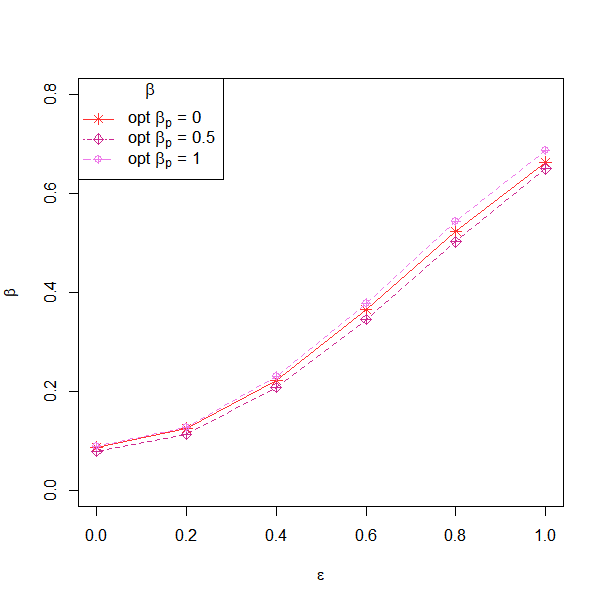}
		\subcaption{$\theta_0$-contaminated cell}
	\end{subfigure}
	\begin{subfigure}{0.5\textwidth}
		\includegraphics[height=8cm, width=8cm]{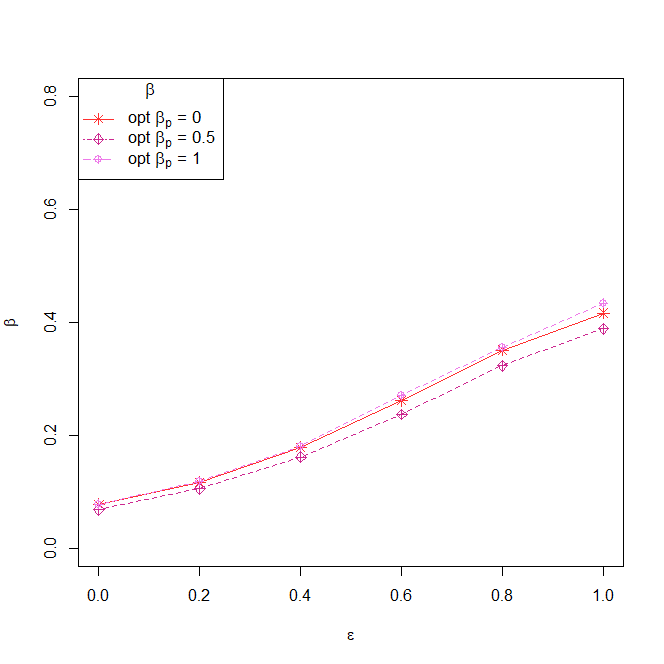}
		\subcaption{$\theta_1$-contaminated cell}
	\end{subfigure}
	\caption{Optimal values of $\beta$ against contamination cell proportion for different pilot estimators based on $R=1000$ repetitions}
	\label{fig:choicebeta}
\end{figure}

\begin{figure}[htb]
	\begin{subfigure}{0.5\textwidth}
		\includegraphics[height=8cm, width=8cm]{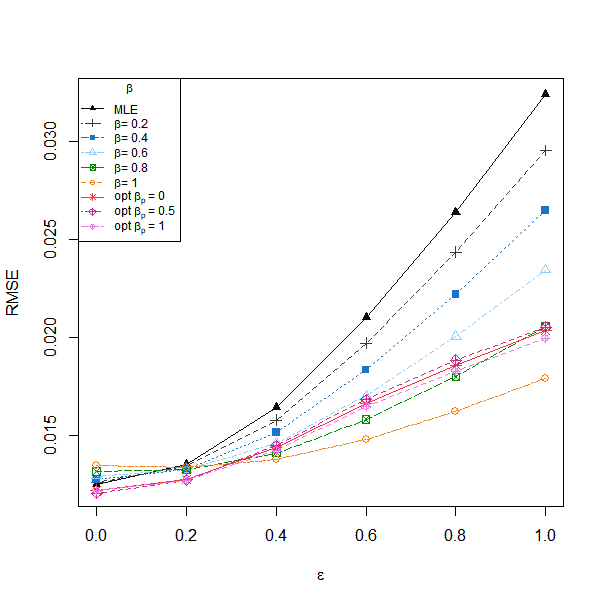}
		\subcaption{$\theta_0$-contaminated cell}
	\end{subfigure}
	\begin{subfigure}{0.5\textwidth}
		\includegraphics[height=8cm, width=8cm]{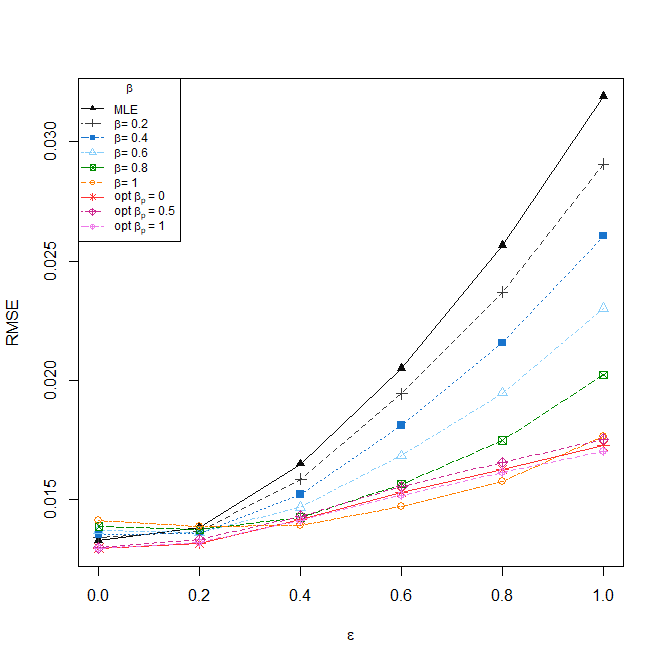}
		\subcaption{$\theta_1$-contaminated cell}
	\end{subfigure}
	\caption{RMSE against contamination cell proportion for different values of the tuning parameter $\beta$ and optimum values obtained with three different pilot estimators based on $R=1000$ repetitions}
	\label{fig:RMSEchoicebeta}
\end{figure}

Figure \ref{fig:choicebeta} shows the optimal values of the tuning parameter $\beta$ against data contamination over $R=1000$ repetitions.
As expected, optimal values of $\beta$ are greater with high contamination rates. Further, the choice of optimal $\beta$ is almost entirely independent of the pilot estimator, and so the presented algorithm does not seem to get affected by this initial choice. Optimal values of the tuning parameter are generally larger when the contamination is introduced on the first parameter. Then, the model is more sensitive to contamination in such direction.
Moreover, in Figure \ref{fig:RMSEchoicebeta}, the RMSE of the resulting (optimal) estimator is compared to the RMSE of the estimators with fixed values of $\beta \in \{0,0.2,0.4,0.6,0.8,1\}$ based on $R=1000$ repetitions. As expected, the data-based method for choosing optimal $\beta$ outperforms any of the methods with a pre-fixed value of $\beta$, since it adapts the tuning parameter value to the amount of contamination present in the data. However, it tends to be slightly conservative and selects insufficiently high values of $\beta$ when a high contamination rate is introduced.
%Since the true distribution underlying the data is unknown, the matrices $\boldsymbol{J}_\beta$ and $\boldsymbol{K}_\beta$ are in turn robustly estimated with the empirical distribution $\boldsymbol{G}_n$ in the forms of  

%Conversely, the proposed algorithm for optimal choice of $\beta$ is sensitive to the sample size. Figure \ref{fig:RMSEchoicebetaN} plots the RMSE of MDPPE with optimal choice of $\beta$ jointly with the RMSE of the MDPDE for fixed values of the tuning parameter against sample size under a $40\%$ of contamination rate. Here  
%
%\begin{figure}[htb]
%	\begin{subfigure}{0.5\textwidth}
%		\includegraphics[height=8cm, width=8cm]{images/choicebetaagaisntN0}
%		\subcaption{$\theta_0$-contaminated cell}
%	\end{subfigure}
%	\begin{subfigure}{0.5\textwidth}
%		\includegraphics[height=8cm, width=8cm]{images/choicebetaagaisntN1}
%		\subcaption{$\theta_1$-contaminated cell}
%	\end{subfigure}
%	\caption{RMSE against contamination cell proportion for different values of the tuning parameter $\beta$ and optimum values obtained with three different pilot estimators}
%	\label{fig:RMSEchoicebetaN}
%\end{figure}
%Discussion between sample size and optimal choice of beta
\section{Data analysis } \label{sec:realdata}

In this section, we discuss two real-life applications of the MDPDE for the SSALT model, developed in the preceding sections.

\subsection{Electronic components data }
The first example was studied by Wang and Fei (2003) to get the reliability indices of a kind of electronic components at the normal temperature of $x_0 = 25^{\circ} C.$ $N=100$ items from a batch of products were randomly selected for a simple SSALT $(k=2)$, with two stress levels $x_1=100^{\circ} C$ and $x_2= 150^{\circ} C.$ In the original experiment, the stress level rises when 30 products had failed and the test continues until 20 more products had failed, obtaining a total of $50$ failures. Their failure times are as follows:
\begin{itemize}
	\item Failure times at the first stress level $x_1$: 32, 54, 59, 86, 117, 123, 213, 267, 268, 273, 299, 311, 321, 333, 339, 386, 408, 422, 435, 437, 476, 518, 570, 632, 666, 697, 796, 854, 858, 910.
	\item Failure times at the second stress level, $x_2$: 16, 19, 21, 36, 37, 63, 70, 75, 83, 95, 100, 106, 110, 113, 116, 135, 136, 149, 172, 186.
\end{itemize}

For illustrating the performance of the MPDPE for the SSALT model with one-shot devices, we will assume that we only know how many devices had failed before certain pre-specified inspection times,  $t= 270,430,600,910,975,1015,1040,1096.$ Additionally, the time of stress change, $\tau_1,$ is pre-fixed at $t= 910.$
% and the vector of stress level is standardized, yielding a stress vector with components $$s_i= \frac{x_i-x_0}{x_k-x_0}, i=1,2.$$

Table \ref{table:estimatedelectronic} shows the estimated model parameters with different values of the tuning parameter $\beta$, along with an approximate CI constructed from (\ref{eq:CI}). The last row of the table contains the estimates with the optimal value of $\beta$ obtained with the algorithm presented in Section \ref{sec:choicebeta}. The algorithm is initialized with the pilot estimate $\widehat{\boldsymbol{\theta}}^{0.5}$ and the optimum value of $\beta$ is reached at $\beta = 0.027$, implying moderately low contamination in the data.
Further, to fairly analyze the role of each of the parameters in the model, we present the logarithm of the first parameter, $\theta_0$, so both model parameters are reported on the same scale. The mean lifetime of the electronic component at low temperatures is appreciably high, and increasing a degree on the temperature multiplies the lifetime of the device, according to all estimators, in approximately 0.97 times. Consequently, the increase of the temperature from $x_1=100^\circ$ to $x_2=150^\circ$ shortens the lifetime by more than 0.22 times.
Robust methods tend to estimate with a higher value the first parameter, and consequently decrease the estimate of the second one.
Conversely, the variance of the estimator increases with $\beta$, producing wider intervals.

\begin{table}[ht]
	\centering
	\caption{Estimated parameters for the electronic components data for different values of $\beta$}
	\label{table:estimatedelectronic}
	\begin{tabular}{ccccc}
		\hline
		$\beta$ & $\log(\widehat{\theta}_0)$  & IC($\log(\theta_0)$) & $\widehat{\theta}_1$ ($\times 10^2$) & IC($\theta_1$) ($\times 10^2$) \\ 
		 \hline
		MLE & -10.857 & [-12.243, -9.470] & 3.021 & [1.887, 4.155] \\ 
		  0.2 & -10.842 & [-12.236, -9.448] & 3.003 & [1.862, 4.143] \\ 
		  0.4 & -10.833 & [-12.236, -9.429] & 2.992 & [1.843, 4.141] \\ 
	  0.6 & -10.827 & [-12.243, -9.411] & 2.986 & [1.827, 4.146] \\ 
		  0.8 & -10.830 & [-12.260, -9.399] & 2.989 & [1.819, 4.160] \\ 
		  1 & -10.837 & [-12.284, -9.389] & 2.996 &[1.813, 4.180] \\ 
		\textbf{0.027}$^\ast$ & \textbf{-10.856} & \textbf{[-12.243, -9.468]} & \textbf{3.019} & \textbf{[1.884, 4.154]} \\
		\hline
	\end{tabular}
\end{table}

Table \ref{table:meanelectronic} shows the mean lifetime (in hours) of the electronic components under different (constant) stress level, and their associated direct and transformed CI. Under normal operating conditions ($x_0=25$), the devices are expected to last for more than 6 hours, but their lifetime gets severely decreased when exposed to very high temperatures, which allows to infer about the reliability in a short period of time. It is interesting to note that direct CI of the mean lifetime under normal operating conditions gets truncated due to the positivity constraint. On the other hand, transformed CI is wider, and the right end point is quite far from than the the one obtained with direct CIs. This difference gets reduced at increased temperatures.

\begin{table}[ht]
	\centering
	\caption{Estimated mean lifetime and asymptotic (direct and transformed) confidence intervals (in hours) of the electronic components under three constant temperatures. }
	\label{table:meanelectronic}
	\begin{tabular}{ccccc}
		\hline
		& Mean lifetime & Direct CI & Transformed CI \\ 
		\hline
		\multicolumn{4}{c}{$x_0=25$}\\
		\hline
		MLE & 6.772 & [0, 14.290] & [2.231, 20.553] \\ 
		0.2 & 6.704 & [0, 14.182] & [2.197, 20.455] \\ 
		0.4 & 6.660 & [0, 14.142] & [2.166, 20.480] \\ 
		0.6 & 6.630 & [0, 14.145] & [2.135, 20.594] \\ 
		0.8 & 6.643 & [0, 14.250] & [2.114, 20.878] \\ 
		1 & 6.679 & [0, 14.417] & [2.096, 21.278] \\ 
		\textbf{0.027$^\ast$} & \textbf{6.768 }& \textbf{[0, 14.589]} & $^\ast$ \\ 
		\hline
		\multicolumn{4}{c}{$x_1=100$}\\
		\hline
		MLE & 0.702 & [0.452, 0.953] & [0.492, 1.004] \\ 
		 0.2 & 0.705 & [0.453, 0.957] & [0.493, 1.008] \\ 
		 0.4 & 0.706 & [0.452, 0.960] & [0.493, 1.011] \\ 
		 0.6 & 0.706 & [0.451, 0.961] & [0.492, 1.013] \\ 
		 0.8 & 0.706 & [0.449, 0.962] & [0.491, 1.015] \\ 
		 1 & 0.706 & [0.447, 0.965] & [0.489, 1.018] \\ 
		\textbf{ 0.027$^\ast$ }& \textbf{0.703} & \textbf{[0.446, 0.961]} & \textbf{[0.488, 1.014]} \\ 
		\hline
		\multicolumn{4}{c}{$x_2=150$}\\
		\hline
		MLE & 0.155 & [0.087, 0.223] & [0.100, 0.241] \\ 
		 0.2 & 0.157 & [0.087, 0.227] & [0.101, 0.245] \\ 
		 0.4.& 0.158 & [0.088, 0.229] & [0.101, 0.247] \\ 
		 0.6 & 0.159 & [0.088, 0.230] & [0.101, 0.248] \\ 
		 0.8 & 0.158 & [0.087, 0.230] & [0.101, 0.248] \\ 
		 1 & 0.158 & [0.087, 0.229] & [0.100, 0.248] \\ 
		 \textbf{0.027$^\ast$} &\textbf{ 0.155 }&\textbf{ [0.086, 0.225]} & \textbf{[0.099, 0.243]} \\ 
		\hline
	\end{tabular}
\end{table}

On the other hand, one may be interested in estimating the reliability of the devices when it is exposed to different constant temperatures. We fix a ``mission time'' at $t=600s$ and we report the estimated reliabilities and CIs under different stress levels in Table \ref{table:reliabilityelectronic}. Again,  direct CIs  had to be truncated under normal operating conditions so as to remain within the interval $(0,1).$ As expected, the reliability of the devices decreases when increasing the stress level, and here all estimates remain  close for all values of the tuning parameter.
\begin{table}[ht]
	\centering
	\caption{Estimated reliability at $t=600s$ and asymptotic (direct and transformed) confidence intervals of the electronic components under three constant temperatures.}
	\label{table:reliabilityelectronic}
	\begin{tabular}{cccc}
		\hline
		& $\widehat{S}(600)$ & Direct CI & Transformed CI \\ 
			\hline
		\multicolumn{4}{c}{$x_0=25$}\\
		\hline
		MLE & 0.976 & [0.949, 1.00] & [0.929, 0.992] \\ 
		 0.2 & 0.975 & [0.948, 1.00] & [0.928, 0.992] \\ 
		 0.4 & 0.975 & [0.948, 1.00] & [0.927, 0.992] \\ 
		 0.6 & 0.975 & [0.947, 1.00] & [0.926, 0.992] \\ 
		 0.8 & 0.975 & [0.947, 1.00] &[0.925, 0.992] \\ 
		 1 & 0.975 & [0.947, 1.00] & [0.924, 0.992] \\ 
		 \textbf{0.027}$^\ast$ & \textbf{0.976} & \textbf{[0.948, 1.00]} & \textbf{[0.926, 0.992]} \\ 
		\hline
		\multicolumn{4}{c}{$x_1=100$}\\
		\hline
		MLE & 0.789 & [0.722, 0.856] & 0.714, 0.848] \\ 
		 0.2 & 0.790 & [0.723, 0.856] & 0.715, 0.849] \\ 
		 0.4 & 0.790 & [0.723, 0.857] & 0.715, 0.849] \\ 
		 0. & 0.790 & [0.722, 0.857] & 0.715, 0.849] \\ 
		 0.8 & 0.790 & [0.722, 0.857] & 0.714, 0.850] \\ 
		 1 & 0.790 & [0.721, 0.858] & 0.713, 0.850] \\ 
		 \textbf{0.027$^\ast$}& \textbf{0.789 }& \textbf{[0.721, 0.857]}&\textbf{[0.713, 0.849]} \\ 
		\hline
		\multicolumn{4}{c}{$x_2=150$}\\
		\hline
		MLE & 0.341 & [0.180, 0.503] & [0.202, 0.516] \\ 
		 0.2 & 0.346 & [0.183, 0.509] & [0.205, 0.521] \\ 
		 0.4 & 0.349 & [0.185, 0.513] & [0.206, 0.524] \\ 
		 0.6 & 0.350 & [0.185, 0.514] & [0.207, 0.526] \\ 
		 0.8 & 0.349 & [0.184, 0.514] & [0.206, 0.526] \\ 
		 1 & 0.348 & [0.182, 0.514] & [0.204, 0.526] \\ 
		 \textbf{0.027$^\ast$} &\textbf{ 0.342} & \textbf{[0.178, 0.507]} & \textbf{[0.200, 0.520]} \\ 
		\hline
	\end{tabular}
\end{table}

Finally, one may be interested in determining the time at which  $10\%$ of the devices are expected to fail, under different (constant) temperatures. Table \ref{table:quantileelectronic} presents the estimated $0.9-$quantiles of the lifetime distribution (or equivalently the $0.1-$quantiles of the reliability). Here, the direct CI of the estimated quantiles under normal operating conditions are again truncated, demonstrating again the drawback of the direct method, while transformed CI provides a good  alternative for such intervals without the problem of constraints.
\begin{table}[ht]
	\centering
	\caption{Estimated $90\%$ quantile  and asymptotic (direct and transformed) confidence intervals (in seconds) of the electronic components under three constant temperatures.}
	\label{table:quantileelectronic}
	\begin{tabular}{cccc}
		\hline
		& $\widehat{Q}_{0.9}$ & Direct CI & Transformed CI \\ 
			\hline
		\multicolumn{4}{c}{$x_0=25$}\\
				\hline
				MLE & 2568.46 & [0, 5420.08] & [846.25, 7795.56] \\ 
				0.2 & 2542.69 & [0, 5379.29] & [833.30, 7758.67] \\ 
				0.4 & 2526.14 & [0, 5363.87] & [821.48, 7768.17] \\ 
				0.6 & 2514.90 & [0, 5365.10] & [809.70, 7811.19] \\ 
				0.8 & 2519.62 & [0, 5405.00] & [801.67, 7919.06] \\ 
				1 & 2533.16 & [0, 5468.50] & [795.09, 8070.66] \\ 
				\textbf{0.027$^\ast$} & \textbf{2567.25} & \textbf{[0, 5533.47]} & \textbf{[808.50, 8151.84]} \\ 
					\hline
				\multicolumn{4}{c}{$x_1=100$}\\
				\hline
				MLE & 266.45 & [171.42, 361.49] & [186.52, 380.65] \\ 
				0.2 & 267.47 & [171.79, 363.14] & [187.03, 382.49] \\ 
				0.4 & 267.82 & [171.62, 364.03] & [187.00, 383.57] \\ 
				0.6 & 267.77 & [171.07, 364.46] & [186.61, 384.22] \\ 
				0.8 & 267.71 & [170.39, 365.04] & [186.12, 385.08] \\ 
				1 & 267.73 & [169.63, 365.84] & [185.60, 386.22] \\ 
				\textbf{0.027$^\ast$}& \textbf{266.75} & \textbf{[169.13, 364.37]} & \textbf{[185.00, 384.62]} \\ 
					\hline
				\multicolumn{4}{c}{$x_2=150$}\\
					\hline
				MLE & 58.83 & [32.89, 84.77] & [37.85, 91.43] \\ 
				0.2 & 59.60 & [33.16, 86.04] & [38.24, 92.88] \\ 
				0.4 & 59.99 & [33.23, 86.76] & [38.41, 93.72] \\ 
				0.6 & 60.15 & [33.19, 87.11] & [38.42, 94.17] \\ 
				0.8 & 60.06 & [33.04, 87.07] & [38.30, 94.17] \\ 
				1 & 59.85 & [32.83, 86.87] & [38.11, 94.00] \\ 
				\textbf{0.027$^\ast$} & \textbf{58.96} & \textbf{[32.50, 85.42]} & \textbf{[37.64, 92.35]} \\ 
				\hline
%		 \textbf{0.027$^\ast$} & \textbf{20.276} & \textbf{[0, 43.704]} &\textbf{ [6.386, 64.384]} \\
%		 	\hline
%		 \multicolumn{4}{c}{$x_0=100$}\\
%		 \hline 
%		MLE & 2.104 & [1.354, 2.855] & [1.473, 3.006] \\ 
%		 0.2 & 2.112 & [1.357, 2.868] & [1.477, 3.021] \\ 
%		 0.4 & 2.115 & [1.355, 2.875] & [1.477, 3.029] \\ 
%		 0.6 & 2.115 & [1.351, 2.879] & [1.474, 3.035] \\ 
%		 0.8 & 2.114 & [1.346, 2.883] & [1.470, 3.041] \\ 
%		 1 & 2.115 & [1.340, 2.889] & [1.465, 3.050] \\ 
%		 \textbf{0.027$^\ast$} & \textbf{2.107} & \textbf{[1.336, 2.878]} & \textbf{[1.461, 3.038]} \\ 
%		 	\hline
%		 \multicolumn{4}{c}{$x_0=150$}\\
%		 \hline
%		MLE & 0.465 & [0.260, 0.670] & [0.299, 0.722] \\ 
%		 0.2 & 0.471 & [0.262, 0.680] & [0.302, 0.734] \\ 
%		 0.4 & 0.474 & [0.262, 0.685] & [0.303, 0.740] \\ 
%		 0.6 & 0.475 & [0.262, 0.688] & [0.303, 0.744] \\ 
%		 0.8 & 0.474 & [0.261, 0.688] & [0.303, 0.744] \\ 
%		 1 & 0.473 & [0.259, 0.686] & [0.301, 0.742] \\ 
%		 \textbf{0.027$^\ast$ }& \textbf{0.466} & \textbf{[0.257, 0.675]} & \textbf{[0.297, 0.729]} \\ 
%		\hline
	\end{tabular}
\end{table}

\subsection{Light bulbs data}

Zhu (2010) conducted an accelerated life testing experiment in the Quality and Reliability Engineering Laboratory of the Industrial and Systems Engineering Department of Rutgers University so as to examine the reliability of light bulbs. Two sets of 32 miniature light bulbs were placed in a temperature and humidity chamber where humidity was held constant, and the long term failure due to bulb filament fatigue was then studied.
%The design working conditions of this light bulb are:
%voltage: 2 Volts,
%current: 0.06 amps.
When the switch was turned on, full current suddenly flowed to the filament at the speed of light. This sudden massive vibration caused the filament to wildly bounce causing fatigue behaviour of the filament which resulted in breakage of the filament. Long Term Failure occurred when the filament eventually become so fatigued that its electrical resistance increased to the point that current would not flow.
Each light bulb was connected with a resistor, across which Voltage was measured to monitor the status of the light bulbs.
%National Instruments LabView software is used to develop the application for the continuous monitoring of the status of the test units.
%The failure time data are automatically saved 
Normal operating condition of the light bulbs is 2V. To carry out the SSALT, they applied 2.25V for 96hr and then increased the voltage to 2.44V.  The step-voltage test got stopped at 140hr. Failure times during the experiment  are as follows:
\begin{center}
	     12.07, 19.5, 22.1, 23.11, 24, 25.1, 26.9, 36.64, 44.1, 46.3, 54, 58.09, 64.17, 72.25, 86.9, 90.09, 91.22, 102.1, 105.1, 109.2, 114.4, 117.9, 121.9, 122.5, 123.6, 126.5, 130.1, 14 17.95, 24, 26.46, 26.58, 28.06, 34, 36.13, 40.85, 41.11 42.63, 52.51, 62.68, 73.13, 83.63, 91.56, 94.38, 97.71, 101.53, 105.11 112.11, 119.58 ,120.2, 126.95, 129.25, 136.31.
\end{center}
The remaining 11 light bulbs continued to provide light when the experiment was terminated. To illustrate the performance of the MDPDE of the SSALT model, we transformed the  collected data into one-shot devices data with inspection times $t=25,50,96,110,120,140$. Table \ref{table:lightbulb} shows the estimated values of the model parameters with different values of the tuning parameter $\beta.$ Applying the data-based choice of $\beta$ described in Section \ref{sec:choicebeta}, the optimum value is approximately $\beta = 0.12,$ thus showing slightly  higher contamination to be present in this data than in the last electronic component example. Results for this optimum value are presented in the last row of  Table \ref{table:lightbulb}. As in the previous example, we report the values of $\log(\widehat{\theta}_0)$ so both model parameters are in same scale. 
Unlike in the previous example, now robust estimators  give a higher value to the second parameter and decrease the value of the first one. 

%\begin{table}[ht]
%	\centering
%	\caption{Estimated values and approximate confidence intervals for the model parameters with light bulb data.}
%	\label{table:lightbulb}
%	\begin{tabular}{ccc|cc}
%		\hline
%		$\beta$ & $\theta_0$ ($\times 10^{-3}$) & IC($\theta_0$) ($\times 10^{-3}$) & $\theta_1$ & IC($\theta_1$)  \\ 
%		\hline
%		0 & 2.067 & [1.022, 3.112] & 2.325 & [1.651, 2.999] \\ 
%		0.2 & 2.054 & [1.021, 3.112] & 2.335 & [1.651, 3.000] \\ 
%		0.4 & 2.043 & [1.019, 3.114] & 2.343 & [1.649, 3.001] \\ 
%		0.6 & 2.026 & [1.016, 3.117] & 2.356 & [1.648, 3.003] \\ 
%		0.8 & 2.010 & [1.013, 3.121] & 2.368 & [1.646, 3.005] \\ 
%		1 & 1.991 & [1.009, 3.124] & 2.381 & [1.644, 3.007] \\ 
%		\hline
%	\end{tabular}
%\end{table}

\begin{table}[ht]
	\centering
		\caption{Estimated parameters for the light bulb data for different values of $\beta.$}
	\label{table:lightbulb}
	\begin{tabular}{ccccc}
	\hline
	$\beta$ & $\log(\widehat{\theta}_0)$  & IC($\log(\theta_0)$) & $\widehat{\theta_1}$ & IC($\theta_1$)  \\ 
	\hline
	MLE & -10.727 & [-11.718, -9.736] & 5.285 & [2.282, 8.287] \\ 
	0.2 & -10.734 & [-11.725, -9.743] & 5.308 & [2.305, 8.310] \\ 
	0.4 & -10.739 & [-11.731, -9.746] & 5.326 & [2.320, 8.332] \\ 
	0.6 & -10.747 & [-11.742, -9.752] & 5.354 & [2.343, 8.364] \\ 
	0.8 & -10.755 & [-11.753, -9.757] & 5.381 & [2.364, 8.398] \\ 
	1 & -10.764 & [-11.766, -9.763] & 5.411 & [2.387, 8.434] \\ 
	\textbf{0.12$^\ast$} & \textbf{-10.729} &\textbf{ [-11.720, -9.738]} &\textbf{ 5.293} &\textbf{ [2.290, 8.295]} \\ 
		\hline
	\end{tabular}
\end{table}

Table \ref{table:meanbulb} shows the estimated mean lifetime and their corresponding direct and transformed CIs, with different values of $\beta,$  under three stress levels: $x_0 = 2$V corresponding to working condition, and two stress levels at which devices were subjected during the experiment, $x_1 = 2.25$V and $x_2= 2.44$V. It is striking that robust methods provide larger mean lifetime under normal operating condition ($x_0=2$V) than the MLE, but shorter lifetimes under a higher stress level ($x_2=2.44$V). 
Furthermore, direct and transformed CIs of the mean lifetime markedly differ under working condition, but that difference gets reduced when the voltage gets increased.
\begin{table}[ht]
		\centering
		\caption{Estimated mean lifetime (in minutes) and asymptotic confidence intervals for the light bulbs under different voltages. }
		\label{table:meanbulb}
	\begin{tabular}{cccc}
		\hline
		& Mean lifetime & Direct CI & Transformed CI \\ 
		\hline
		\multicolumn{4}{c}{$x_0=2$}\\
		\hline
		MLE & 483.84 & [4.48, 963.19] & [179.65, 1303.08] \\ 
		0.2 & 486.96 & [4.33, 969.59] & [180.74, 1311.99] \\ 
		0.4 & 489.43 & [3.56, 975.31] & [181.37, 1320.78] \\ 
		0.6 & 493.47 & [2.42, 984.51] & [182.43, 1334.82] \\ 
		0.8 & 497.58 & [0.96, 994.20] & [183.40, 1349.95] \\ 
		1 & 502.22 & [0, 1005.14] & [184.50, 1367.09] \\ 
		\textbf{0.12$^\ast$} & \textbf{484.82} & \textbf{[0, 970.90]} &\textbf{ [177.89, 1321.32]} \\ 
		 	\hline
		\multicolumn{4}{c}{$x_0=2.25$}\\
		\hline
		MLE & 129.10 & [85.30, 172.90] & [91.96, 181.25] \\ 
		0.2 & 129.19 & [85.33,173.06] & [92.00, 181.42]\\ 
		0.4 & 129.25 & [85.29, 173.21] & [91.99, 181.61] \\ 
		0.6 & 129.41 & [85.28, 173.54] & [92.02, 182.00] \\ 
		0.8 & 129.61 & [85.27, 173.95] & [92.06, 182.48] \\ 
		1 & 129.85 & [85.25, 174.44] & [92.10, 183.06] \\ 
		\textbf{0.12$^\ast$} & \textbf{129.11} & \textbf{[84.86, 173.35]} &\textbf{ [91.64, 181.88]} \\
		 \hline
		\multicolumn{4}{c}{$x_0=2.44$}\\
		\hline
		MLE & 47.30 & [25.60, 69.00] & [29.90, 74.83] \\ 
		0.2 & 47.13 & [25.53, 68.72] & [29.80, 74.52] \\ 
		0.4 & 46.99 & [25.46, 68.52] & [29.72, 74.30] \\ 
		0.6 & 46.80 & [25.36, 68.23] & [29.60, 73.98] \\ 
		0.8 & 46.63 & [25.27, 67.98] & [29.49, 73.70] \\ 
		1 & 46.45 & [25.19, 67.71] & [29.39, 73.41] \\ 
		\textbf{0.12$^\ast$} & \textbf{47.23} & \textbf{[25.46, 69.01]} &\textbf{ [29.79, 74.90]} \\ 
		\hline
	\end{tabular}
\end{table}
%\begin{table}[ht]
%	\centering
%	\caption{Estimated mean lifetime and approximate confidence intervals for the light bulbs under different stress levels. }
%	\label{table:meanbulb}
%	\begin{tabularx}{0.9\textwidth}{l X c| X c| X c|}%{ccccccc}
%		& \multicolumn{2}{c}{$x_0 = 2$V} & \multicolumn{2}{c}{$x_1 = 2.25$V} & \multicolumn{2}{c}{$x_2 = 2.44$V}\\
%		\hline
%		$\beta$ & mean lifetime & IC & mean lifetime & IC & mean lifetime & IC \\ 
%		\hline
%		0 & 483.84 & [239.26, 728.41] & 129.10 & [106.75, 151.45] & 47.30 & [36.23, 58.37] \\ 
%		0.2 & 486.96 & [240.71, 733.21] & 129.19 & [106.81, 151.57] & 47.13 & [36.11, 58.15] \\ 
%		0.4 & 489.43 & [241.53, 737.33] & 129.25 & [106.83, 151.68] & 46.99 & [36.00, 57.97] \\ 
%		0.6 & 493.47 & [242.93, 744.00] & 129.41 & [106.90, 151.93] & 46.80 & [35.86, 57.73] \\ 
%		0.8 & 497.58 & [244.20, 750.96] & 129.61 & [106.98, 152.23] & 46.63 & [35.73, 57.52] \\ 
%		1 & 502.22 & [245.62, 758.82] & 129.85 & [107.10, 152.60] & 46.45 & [35.60, 57.30] \\ 
%		\hline
%	\end{tabularx}
%\end{table}
Next, Table \ref{table:reliabilitylightbulb} shows the estimated reliability at mission time $t=50s$ under constant voltage, with different values of $\beta.$ The reliability of the light bulbs at working condition is sufficiently high, exceeding a $90\%$ reliability, but gets radically decreased at the highest voltage $x_2=2.44.$
\begin{table}[ht]
	\centering
	\caption{Estimated reliability at $t=50$ seconds and asymptotic (direct and transformed) confidence intervals for the light bulbs under three constant voltages.}
	\label{table:reliabilitylightbulb}
	\begin{tabular}{cccc}
		& $\widehat{S}(50)$ & Direct CI & Transformed CI \\ 
		\hline
		\multicolumn{4}{c}{$x_0=2$}\\
		\hline
		MLE & 0.902 & [0.809, 0.994]  & [0.764, 0.963] \\ 
		 0.2 & 0.902 & [0.811, 0.994]  & [0.765, 0.963] \\ 
		 0.4 & 0.903 & [0.811, 0.994]  & [0.766, 0.964] \\ 
		 0.6 & 0.904 & [0.813, 0.995]   & [0.767, 0.964] \\ 
		 0.8 & 0.904 & [0.814, 0.995]  & [0.768, 0.964] \\ 
		 1 & 0.905 & [0.815, 0.995]  & [0.769, 0.965] \\ 
		 \textbf{0.12$^\ast$} & \textbf{0.902} & \textbf{[0.809, 0.995] } &\textbf{ [0.762, 0.964]} \\
		 \hline
		 \multicolumn{4}{c}{$x_0=2.25$}\\
		 \hline 
		MLE & 0.679 & [0.590, 0.768]  & [0.584, 0.761] \\ 
		 0.2 & 0.679 & [0.590, 0.768] & [0.584, 0.761] \\ 
		 0.4 & 0.679 & [0.590, 0.769] & [0.584, 0.761] \\ 
		 0.6 & 0.680 & [0.590, 0.769]  & [0.584, 0.762] \\ 
		 0.8 & 0.680 & [0.590, 0.770]  & [0.584, 0.762] \\ 
		 1 & 0.680 & [0.590, 0.770]  & [0.585, 0.763] \\ 
		 \textbf{0.12$^\ast$} & \textbf{0.679} & \textbf{[0.589, 0.769]}  & \textbf{[0.583, 0.762]} \\
		 \hline
		 \multicolumn{4}{c}{$x_0=2.44$}\\
		 \hline 
		MLE & 0.347 & [0.179, 0.516] & [0.202, 0.528] \\ 
		 0.2 & 0.346 & [0.178, 0.514] & [0.201, 0.527] \\ 
		 0.4 & 0.345 & [0.177, 0.513]& [0.200, 0.526] \\ 
		 0.6 & 0.344 & [0.175, 0.512]  & [0.199, 0.524] \\ 
		 0.8 & 0.342 & [0.174, 0.510]  & [0.198, 0.523] \\ 
		 1 & 0.341 & [0.173, 0.509]  & [0.197, 0.522] \\ 
		 \textbf{0.12$^\ast$} & \textbf{0.347} &\textbf{ [0.178, 0.516]}  & \textbf{[0.201, 0.529]} \\ 
		\hline
	\end{tabular}
\end{table}

Finally,Table \ref{table:quantilelightbulb} presents the estimated $0.9$-quantile of the lifetime distribution. At that times, $10\%$ of light bulbs are expected to fail if they are subjected to constant voltage. Under normal operating condition $x_0 = 2,$  $10\%$ of the light bulbs are expected to fail by (approximately) $52s.$ Higher quantiles are predicted with  low values of $\beta$, including the MLE, but when increasing voltage to $x_2=2.44,$   the  situation  turns around, and quantiles based on MDPDE with higher values of $\beta$ predict lower times at which $10\%$ of the light bulbs are expected to fail.

\begin{table}[ht]
	\centering
	\caption{Estimated $90\%$ quantile and asymptotic (direct and transformed) confidence intervals (in seconds) of the light bulbs under three constant stress levels.}
	\label{table:quantilelightbulb}
	\begin{tabular}{cccc}
		\hline
	& $\widehat{Q}_{0.9}$ & Direct CI & Transformed CI \\ 
	\hline
	\multicolumn{4}{c}{$x_0=2$}\\
	\hline
		MLE & 50.98 & [0.47, 101.48]  & [18.93, 137.29] \\ 
		0.2 & 51.31 & [0.46, 102.16]& [19.04, 138.23] \\ 
		0.4 & 51.57 & [0.37, 102.76]  & [19.11, 139.16] \\ 
		0.6 & 51.99 & [0.25, 103.73]  & [19.22, 140.64] \\ 
		0.8 & 52.42 & [0.10, 104.75]  & [19.32, 142.23] \\ 
		1 & 52.91 & [0, 105.90] & [19.44, 144.04] \\ 
	\textbf{	0.12$^\ast$ }& \textbf{51.08 }& \textbf{[0, 102.29]} &\textbf{ [18.74, 139.22]} \\ 
		\hline
	\multicolumn{4}{c}{$x_1=2.25$}\\
	\hline
		MLE & 13.60 & [8.99, 18.22] &  [9.69, 19.10] \\ 
		0.2.& 13.61 & [8.99, 18.23] &  [9.69, 19.11] \\ 
		0.4.& 13.62 & [8.99, 18.25] &  [9.69, 19.13] \\ 
		0.6 & 13.64 & [8.99, 18.28] & [9.70, 19.18] \\ 
		0.8 & 13.66 & [8.98, 18.33] &  [9.70, 19.23] \\ 
		1 & 13.68 & [8.98, 18.38] & [9.70, 19.29] \\ 
	\textbf{	0.12$^\ast$} &\textbf{ 13.60} & \textbf{[8.94, 18.26]} & \textbf{ [9.66, 19.16]} \\ 
		\hline
	\multicolumn{4}{c}{$x_2=2.44$}\\
	\hline
		MLE & 4.98 & [2.70, 7.27] & [3.15, 7.88] \\ 
		0.2 & 4.97 & [2.69, 7.24] & [3.14, 7.85] \\ 
		0.4 & 4.95 & [2.68, 7.22] & [3.13, 7.83] \\ 
		0.6 & 4.93 & [2.67, 7.19] &  [3.12, 7.79] \\ 
		0.8 & 4.91 & [2.66, 7.16] &[3.11, 7.77] \\ 
		1 & 4.89 & [2.65, 7.13] &  [3.10, 7.73] \\ 
		\textbf{0.12$^\ast$} & \textbf{4.98} &\textbf{ [2.68, 7.27]}  & \textbf{[3.14, 7.89]} \\ 
		\hline
	\end{tabular}
\end{table}

\section{Concluding remarks \label{sec:concluding} }

In this paper, we have developed robust estimation methods and test procedures for non-destructive one-shot devices under the SSALT model with exponential lifetimes.
The proposed MDPDEs, indexed by a tuning parameter $\beta$ controlling the trade-off between efficiency and robustness, generalize the classical likelihood approach to a wider family, including the MLE for $\beta=0$. MDPDEs are consistent, asymptotically normal and also enjoy robustness properties for positive values of the tuning parameter; they offer a competitive and robust alternative to the classical estimators based on MLEs. 
Additionally, we have presented a data-based criterion for choosing an optimal value of the tuning parameter, $\beta,$ that does not depend on any initial (pilot) estimator.
% and provides
Through MDPDEs, point estimation and direct and transformed CIs of some lifetime characteristics of interest, such as the reliability at certain mission times, distribution quantiles and mean lifetimes, have been proposed.

Further, robust Z-type test statistics based on the MDPDEs have been developed for linear null hypothesis and its level and power functions have been empirically and theoretically studied. Robust estimators and the associated Z-type test statistics perform slightly worse than MLE in low contaminated scenarios, but they exhibit  worthwhile gain in terms of robustness  when contamination is present in a cell through a contamination in any of the model parameters.

Finally, two real data examples have been analyzed to illustrate	all  inferential methods developed here. The data-based choice of the optimal $\beta$ assists us in understanding the level of contamination present in the sample, and then choosing an estimator suitably.

\section*{Fundings}
This work was supported by the Spanish Grants PGC2018-095194-B-100 and FPU/018240  and Natural Sciences and Engineering Research Council of Canada (of the first author) through an Individual Discovery Grant (No. 20013416).

\section*{Data Availability Statement} Data sharing is not applicable to this article as no new data were created or analyzed in this study. The real datasets are publicly available on the corresponding referenced papers.

\section*{Conflicts of Interest}
The authors declare no conflict of interest. The funders had no role in the design of the study; in the collection, analyses, or interpretation of data; in the writing of the manuscript, or in the decision to publish the results.

\section*{ORCID}
Narayanaswamy Balakrishnan{\includegraphics[keepaspectratio,width=0.7em]{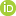}} \hspace{1mm} \href{https://orcid.org/0000-0001-5842-8892}{orcid.org/0000-0001-5842-8892}\\
Elena Castilla  {\includegraphics[keepaspectratio,width=0.7em]{orcidlogo.png}} \hspace{1mm} \href{https://orcid.org/0000-0002-9626-6449}{orcid.org/0000-0002-9626-6449}\\
María Jaenada {\includegraphics[keepaspectratio,width=0.7em]{orcidlogo.png}} \hspace{1mm} \href{https://orcid.org/0000-0002-2874-8286}{orcid.org/0000-0002-2874-8286}\\
Leandro Pardo {\includegraphics[keepaspectratio,width=0.7em]{orcidlogo.png}} \hspace{1mm} \href{https://orcid.org/0000-0003-2005-8245}{orcid.org/0000-0003-2005-8245}

\appendix
\section*{Appendix: Proofs of the main results}

\subsection*{Proof of Result 2}
\begin{proof}
	The MDPDE is defined as the minimizer of the DPD-based loss in (\ref{eq:DPDloss}), and so must satisfy:
	\begin{equation*}
		\frac{\partial d_{\beta}\left( \widehat{\boldsymbol{p}},\boldsymbol{\pi}\left(\boldsymbol{\theta}\right)\right)}{\partial \boldsymbol{\theta}}   = (\beta+1)\sum_{j=1}^{L+1} \left(  \pi_j(\boldsymbol{\theta})^{\beta-1} \left(\pi_j(\boldsymbol{\theta}) -  \widehat{p}_j\right) \frac{\partial \pi_j(\boldsymbol{\theta})}{\partial \boldsymbol{\theta}}  \right) = \boldsymbol{0}_2.
	\end{equation*}
	Next, the derivative of the probability of success $\pi_j(\boldsymbol{\theta})$
	%at each interval $(t_{j-1}, t_j],$ $j=1,...,L$ 
	depends on the stress level at which the device is being tested.
	%Note that we can assume that the devices are submitted to the same stress level throughout the entire interval, which we denoted by $x_i$ for some $i=1,...,k.$
	We denote $x_i$ for the stress level at which the units are tested after the $\tau_i-$th inspection time.
	Taking derivatives in (\ref{eq:th.prob}), we get
	\begin{equation*}
		\frac{\partial\pi_j(\boldsymbol{\theta})}{\partial \boldsymbol{\theta}}
		= \frac{\partial G_T(t_j)}{\partial \boldsymbol{\theta}} -\frac{G_T(t_{j-1})}{\partial \boldsymbol{\theta}}.
	\end{equation*}
	%	Additionally, we define the quantity $a_{i-1}^\ast = \lambda_{i-1} \sum_{l=1}^{i-1}\left(\frac{z_l-z_{l-1}}{\lambda_l}\right)(x_{i-1}-x_l).$
	Upon, using 
	$$ \frac{\partial \lambda_{i}(\boldsymbol{\theta})}{\partial \boldsymbol{\theta}} = \left(\exp(\theta_1 x_i), \theta_0 \exp(\theta_1x_i)x_i\right)^T 
	\hspace{0.3cm} \text{and} \hspace{0.3cm} \frac{\partial a_{i-1}(\boldsymbol{\theta})}{\partial \boldsymbol{\theta}} = (0,  a_{i-1}^\ast)^T
	$$
	with $a_{i-1}^\ast$ defined in (\ref{aast}), we have
	\begin{align*}
		\boldsymbol{z}_j &= \frac{\partial G_T(t_j)}{\partial \boldsymbol{\theta}}= e^{-\lambda_i(t_j+a_{i-1}-\tau_{i-1})}\left(\frac{\partial \lambda_i(\boldsymbol{\theta})}{\partial \boldsymbol{\theta}}(t_j+a_{i-1}-\tau_{i-1}) + \lambda_i(\boldsymbol{\theta})\frac{\partial a_{i-1}(\boldsymbol{\theta})}{\partial \boldsymbol{\theta}}\right)\\
		&= g_T(t_j )\begin{pmatrix}
			\frac{t_j+a_{i-1}-\tau_{i-1}}{\theta_0}\\
			(t_j+a_{i-1}-\tau_{i-1})x_i + a_{i-1}^\ast
		\end{pmatrix} .
	\end{align*}
	Defining the matrix $\boldsymbol{W}$ with rows $\boldsymbol{w}_j =  \boldsymbol{z}_j-\boldsymbol{z}_{j-1},$ we obtain the desired expression.

\end{proof}
\subsection*{Proof of Result 3}
\begin{proof}
	Following Basu et al. (1998), the matrices $\boldsymbol{J}_\beta(\boldsymbol{\theta}_0) $ and $\boldsymbol{K}_\beta(\boldsymbol{\theta}_0) $ are given by
	\begin{align*}
		\boldsymbol{J}_\beta(\boldsymbol{\theta}_0) &= \sum_{j=1}^{L+1} \boldsymbol{u}_j \boldsymbol{u}_j^T \pi_j(\boldsymbol{\theta})^{\beta+1}\\
		\boldsymbol{K}_\beta(\boldsymbol{\theta}_0) &= \sum_{j=1}^{L+1} \boldsymbol{u}_j \boldsymbol{u}_j^T \pi_j(\boldsymbol{\theta}_0)^{2\beta+1} - \left( \sum_{j=1}^{L+1} \boldsymbol{u}_j  \pi_j(\boldsymbol{\theta}_0)^{\beta+1}\right) \left( \sum_{j=1}^{L+1} \boldsymbol{u}_j  \pi_j(\boldsymbol{\theta}_0)^{\beta+1}\right)^T,\\
	\end{align*}
	where 
	$$\boldsymbol{u}_j(\boldsymbol{\theta}) = \frac{\partial \log(\pi_j(\boldsymbol{\theta}))}{\partial \boldsymbol{\theta}} = \frac{1}{\pi_j(\boldsymbol{\theta})}\frac{\partial \pi_j(\boldsymbol{\theta})}{\partial \boldsymbol{\theta}} = \frac{\boldsymbol{w}_j}{\pi_j(\boldsymbol{\theta})}.$$
	Hence, we can write
	\begin{align*}
		\boldsymbol{J}_\beta(\boldsymbol{\theta}_0) &= \sum_{j=1}^{L+1} \boldsymbol{w}_j \boldsymbol{w}_j^T \pi_j(\boldsymbol{\theta}_0)^{\beta-1} = \boldsymbol{W}^T D_{\boldsymbol{\pi}(\boldsymbol{\theta_0})}^{\beta-1} \boldsymbol{W}\\
		\boldsymbol{K}_\beta(\boldsymbol{\theta}_0) &= \sum_{j=1}^{L+1} \boldsymbol{w}_j\boldsymbol{w}_j^T \pi_j(\boldsymbol{\theta}_0)^{2\beta-1} - \left( \sum_{j=1}^{L+1} \boldsymbol{w}_j \pi_j(\boldsymbol{\theta}_0)^{\beta}\right) \left( \sum_{j=1}^{L+1} \boldsymbol{w}_j \pi_j(\boldsymbol{\theta}_0)^{\beta}\right)^T \\ 
		&= \boldsymbol{W}^T \left(D_{\boldsymbol{\pi}(\boldsymbol{\theta}_0)}^{2\beta-1}-\boldsymbol{\pi}(\boldsymbol{\theta}_0)^{\beta}\boldsymbol{\pi}(\boldsymbol{\theta}_0)^{\beta T}\right) \boldsymbol{W}.\\
	\end{align*}
	
\end{proof}

\subsection*{Proof of Result 5}
\begin{proof}
	For notational convenience, let us define $\varepsilon$, $\boldsymbol{\theta}_\varepsilon = \boldsymbol{T}_\beta(\boldsymbol{G}_\varepsilon).$ Here, $\boldsymbol{G}_\varepsilon = (1-\varepsilon)F_{\boldsymbol{\theta}_0} + \varepsilon \Delta_{\boldsymbol{n}}.$ The MDPDE is the minimum of the DPD between $\boldsymbol{\pi}(\boldsymbol{\theta})$ and $\boldsymbol{g}_\varepsilon$, and so must satisfy
	%	\begin{equation}\label{eq:H}
	%		H_{\beta}\left( g,\boldsymbol{\pi}\left(\boldsymbol{\theta}\right)\right)   = \sum_{j=1}^{L+1} \left(\pi_j(\boldsymbol{\theta})^{1+\beta} -\left( 1+\frac{1}{\beta}\right) g_j\pi_j(\boldsymbol{\theta})^{\beta} + \frac{1}{\beta}g_j^{1+\beta} \right)
	%	\end{equation}
	\begin{equation} \label{eq:estimatingH}
		\sum_{j=1}^{L+1} \pi_j(\boldsymbol{\theta}_\varepsilon)^{\beta-1}\left[(\pi_j(\boldsymbol{\theta}_\varepsilon)-\boldsymbol{g}_{\varepsilon,j})\frac{\partial \pi_j(\boldsymbol{\theta}_\varepsilon)}{\partial \boldsymbol{\theta}}\right] = \boldsymbol{0}.
	\end{equation}
	Implicitly differentiating in the estimating equation (\ref{eq:estimatingH}), we obtain
	\begin{align*}
		\sum_{j=1}^{L+1} &(\beta-1)\pi_j(\boldsymbol{\theta}_\varepsilon)^{\beta-2}\frac{\partial \pi_j(\boldsymbol{\theta}_\varepsilon)}{\partial \boldsymbol{\theta}}\frac{\partial \boldsymbol{\theta}_\varepsilon}{\partial \varepsilon}\left[(\pi_j(\boldsymbol{\theta}_\varepsilon)-\boldsymbol{g}_{\varepsilon,j})\frac{\partial \pi_j(\boldsymbol{\theta}_\varepsilon)}{\partial \boldsymbol{\theta}}\right]\\
		& + \pi_j(\boldsymbol{\theta}_\varepsilon)^{\beta-1}\left[\left(\frac{\partial \pi_j(\boldsymbol{\theta}_\varepsilon)}{\partial \boldsymbol{\theta}}\frac{\partial \boldsymbol{\theta}_\varepsilon}{\partial \varepsilon}-\frac{\boldsymbol{g}_{\varepsilon,j}}{\partial \varepsilon} \right) \frac{\partial \pi_j(\boldsymbol{\theta}_\varepsilon)}{\partial \boldsymbol{\theta}} + \left(\pi_j(\boldsymbol{\theta}_\varepsilon)-\boldsymbol{g}_{\varepsilon,j}\right) \frac{\partial^2 \pi_j(\boldsymbol{\theta}_\varepsilon)}{\partial \boldsymbol{\theta}^2} \frac{\partial \boldsymbol{\theta}_\varepsilon}{\partial \varepsilon}\right] = \boldsymbol{0}.
	\end{align*}
	Upon using $\boldsymbol{g}_0 = \boldsymbol{\pi}(\boldsymbol{\theta}_0)$ and evaluating at $\varepsilon = 0$, we get
	\begin{equation*}
		\sum_{j=1}^{L+1} \pi_j(\boldsymbol{\theta}_0)^{\beta-1}\left[\left(\frac{\partial \pi_j(\boldsymbol{\theta}_0)}{\partial \boldsymbol{\theta}}\right)^2\text{IF}\left(\boldsymbol{n}, \boldsymbol{T}_\beta, \boldsymbol{G}\right)-\left(\frac{\partial \pi_j(\boldsymbol{\theta}_0)}{\partial \boldsymbol{\theta}}\right)(-\pi_j(\boldsymbol{\theta}_0) + \Delta_{\boldsymbol{n}})   \right] = \boldsymbol{0}.
	\end{equation*}
	Writing the  obtained equations in matrix form 
	$$ \boldsymbol{W}^T D_{\boldsymbol{\pi}(\boldsymbol{\theta_0})}^{\beta-1} \boldsymbol{W} \cdot\text{IF}\left(\boldsymbol{n}, \boldsymbol{T}_\beta, \boldsymbol{G}\right) - \boldsymbol{W}^T \boldsymbol{D}_{\boldsymbol{\pi}(\boldsymbol{\theta}_0)}^{\beta-1}\left(-\boldsymbol{\pi}(\boldsymbol{\theta}_0)+\Delta_{\boldsymbol{n}} \right)$$
	and solving for $\text{IF}\left(\boldsymbol{n}, \boldsymbol{T}_\beta, \boldsymbol{G}\right),$ we obtain the desired expression.
	
\end{proof}

\subsection*{Proof of the Result 11}
\begin{proof}
	Under the null hypothesis, we have
	$$\boldsymbol{m}^T\widehat{\boldsymbol{\theta}}^\beta - d = \boldsymbol{m}^T(\widehat{\boldsymbol{\theta}}^\beta - \boldsymbol{\theta}_0).$$
	Then, from Result \ref{thm:asymptoticestimator}, we know that
	$$ \sqrt{N}\left(\widehat{\boldsymbol{\theta}}^\beta - \boldsymbol{\theta}_0\right) \rightarrow \mathcal{N}\left(\boldsymbol{0}, \boldsymbol{J}_\beta^{-1}(\boldsymbol{\theta}_0)\boldsymbol{K}_\beta(\boldsymbol{\theta}_0)\boldsymbol{J}_\beta^{-1}(\boldsymbol{\theta}_0)\right)$$
	from which it follows that
	$$ \sqrt{N}\left(\boldsymbol{m}^T\widehat{\boldsymbol{\theta}}^\beta - d\right) \rightarrow \mathcal{N}\left(\boldsymbol{0}, \boldsymbol{m}^T\boldsymbol{J}_\beta^{-1}(\boldsymbol{\theta}_0)\boldsymbol{K}_\beta(\boldsymbol{\theta}_0)\boldsymbol{J}_\beta^{-1}(\boldsymbol{\theta}_0)\boldsymbol{m}\right),$$
	and then transforming it, we obtain %$\sqrt{N}\left(\boldsymbol{m}^T\widehat{\boldsymbol{\theta}}^\beta - d\right)$
	%with the matrix $$\left(\boldsymbol{m}^T\boldsymbol{J}_\beta^{-1}(\widehat{\boldsymbol{\theta}}^\beta)\boldsymbol{K}_\beta(\widehat{\boldsymbol{\theta}}^\beta)\boldsymbol{J}_\beta^{-1}(\widehat{\boldsymbol{\theta}}^\beta)\boldsymbol{m}\right)^{-1/2}$$ 
	$$ \sqrt{N}\left(\boldsymbol{m}^T\boldsymbol{J}_\beta^{-1}(\widehat{\boldsymbol{\theta}}^\beta)\boldsymbol{K}_\beta(\widehat{\boldsymbol{\theta}}^\beta)\boldsymbol{J}_\beta^{-1}(\widehat{\boldsymbol{\theta}}^\beta)\boldsymbol{m}\right)^{-1/2}\left(\boldsymbol{m}^T\widehat{\boldsymbol{\theta}}^\beta - d\right) \rightarrow \mathcal{N}\left(\boldsymbol{0}, \boldsymbol{I}_{r\times r}\right).$$
	Now, as $\widehat{\boldsymbol{\theta}}^\beta$ is a consistent estimator of $\boldsymbol{\theta}_0,$ the stated result follows from Slutsky's theorem.
	
\end{proof}

\subsection*{Proof of the Result 13}
\begin{proof}
	We can rewrite
	$$\boldsymbol{m}^T\widehat{\boldsymbol{\theta}}^\beta - d = \boldsymbol{m}^T\boldsymbol{\theta}_L -d + \boldsymbol{m}^T(\widehat{\boldsymbol{\theta}}^\beta-\boldsymbol{\theta}_L) = \frac{1}{\sqrt{N}}\boldsymbol{m}^T \ell + \boldsymbol{m}^T(\widehat{\boldsymbol{\theta}}^\beta-\boldsymbol{\theta}_L)$$
	and then
	$$\sqrt{N}(\boldsymbol{m}^T\widehat{\boldsymbol{\theta}}^\beta - d) = \boldsymbol{m}^T \ell + \boldsymbol{m}^T\sqrt{N}(\widehat{\boldsymbol{\theta}}^\beta-\boldsymbol{\theta}_L).$$
	But, from Result \ref{thm:asymptoticestimator}, we know that
	$$ \sqrt{N}\left(\widehat{\boldsymbol{\theta}}^\beta - \boldsymbol{\theta}_L\right)  \xrightarrow[L\rightarrow\infty]{L} \mathcal{N}\left(\boldsymbol{0}, \boldsymbol{J}_\beta^{-1}(\boldsymbol{\theta}_L)\boldsymbol{K}_\beta(\boldsymbol{\theta}_L)\boldsymbol{J}_\beta^{-1}(\boldsymbol{\theta}_L)\right).$$
	Therefore,
	$$\sqrt{N}(\boldsymbol{m}^T\widehat{\boldsymbol{\theta}}^\beta - d)   \xrightarrow[L\rightarrow\infty]{L} \mathcal{N}\left( \boldsymbol{m}^T \ell, \boldsymbol{m}^T\boldsymbol{J}_\beta^{-1}(\boldsymbol{\theta}_L)\boldsymbol{K}_\beta(\boldsymbol{\theta}_L)\boldsymbol{J}_\beta^{-1}(\boldsymbol{\theta}_L)\boldsymbol{m}\right)$$
	and so
	$$\frac{\sqrt{N}(\boldsymbol{m}^T\widehat{\boldsymbol{\theta}}^\beta - d)-\boldsymbol{m}^T\ell}{\sqrt{\boldsymbol{m}^T\boldsymbol{J}_\beta^{-1}(\boldsymbol{\theta}_L)\boldsymbol{K}_\beta(\boldsymbol{\theta}_L)\boldsymbol{J}_\beta^{-1}(\boldsymbol{\theta}_L)\boldsymbol{m}}}  \xrightarrow[L\rightarrow\infty]{L} \mathcal{N}\left(0, 1 \right).$$
	As $\widehat{\boldsymbol{\theta}}^\beta \xrightarrow[]{P} \boldsymbol{\theta}_L$,  the stated result follows from Slutsky’s theorem.
	
\end{proof}

\subsection*{Proof of the Result 14}

\begin{proof}
	The power function is the probability of rejection, given the critical region in (\ref{eq:criticalregionZtest}). Thus,
	\begin{align*}
		\beta_N\left(\boldsymbol{\theta}^\ast\right) &= \mathbb{P}\left(|Z_{N}(\widehat{\boldsymbol{\theta}}^\beta)| > z_{\alpha/2} | \boldsymbol{\theta} = \boldsymbol{\theta}^\ast \right)\\
		&= 2\mathbb{P}\left(Z_{N}(\widehat{\boldsymbol{\theta}}^\beta) > z_{\alpha/2} | \boldsymbol{\theta} = \boldsymbol{\theta}^\ast \right)\\
		&= 2 \mathbb{P}\bigg(\sqrt{\frac{N}{\boldsymbol{m}^T\boldsymbol{J}_\beta^{-1}(\widehat{\boldsymbol{\theta}}^\beta)\boldsymbol{K}_\beta(\widehat{\boldsymbol{\theta}}^\beta)\boldsymbol{J}_\beta^{-1}(\widehat{\boldsymbol{\theta}}^\beta)\boldsymbol{m}}} \left(\boldsymbol{m}^T\widehat{\boldsymbol{\theta}}^\beta - \boldsymbol{\theta}^\ast \right) \\
		& \hspace{1cm}> z_{\alpha/2} - \sqrt{\frac{N}{\boldsymbol{m}^T\boldsymbol{J}_\beta^{-1}(\widehat{\boldsymbol{\theta}}^\beta)\boldsymbol{K}_\beta(\widehat{\boldsymbol{\theta}}^\beta)\boldsymbol{J}_\beta^{-1}(\widehat{\boldsymbol{\theta}}^\beta)\boldsymbol{m}}} \left( \boldsymbol{\theta}^\ast -d \right)\bigg).
	\end{align*}
	As $\widehat{\boldsymbol{\theta}}^\beta \xrightarrow[]{P} \boldsymbol{\theta}^\ast$ and
	$ \sqrt{N}\left(\widehat{\boldsymbol{\theta}}^\beta - \boldsymbol{\theta}^\ast\right)  \xrightarrow[L\rightarrow\infty]{L} \mathcal{N}\left(\boldsymbol{0}, \boldsymbol{J}_\beta^{-1}(\boldsymbol{\theta}_L)\boldsymbol{K}_\beta(\boldsymbol{\theta}_L)\boldsymbol{J}_\beta^{-1}(\boldsymbol{\theta}_L)\right),$
	the result follows from Slutsky’s theorem.
	
\end{proof}

\end{document}